\documentclass[final]{agujournal2019}
\usepackage{url} \usepackage{lineno}
\usepackage[inline]{trackchanges}
\usepackage{soul}
\usepackage{booktabs}
\usepackage{float}  
\usepackage{amsmath,amssymb}

\journalname{Machine Learning and Computation}

\begin{document}

\title{GLONET: Mercator's end-to-end neural Global Ocean forecasting system}
\authors{Anass El Aouni, Quentin Gaudel, Charly Regnier, Simon Van Gennip,
Olivier Le Galloudec, Marie Drevillon, Yann Drillet, Jean-Michel Lellouche}

\affiliation{}{Mercator Ocean International, Toulouse, France.}

\correspondingauthor{Anass El Aouni}{aelaouni@mercator-ocean.eu}

\begin{keypoints}

\item GLONET combines physics-based principles with neural networks to effectively capture local-global ocean interactions. 

\item A series of comprehensive validation metrics is proposed, specifically tailored for neural network-based ocean forecasting systems.

\item GLONET's experimental daily forecasts are accessible through the European Digital Twin Ocean platform EDITO.
\end{keypoints}

\begin{abstract}
Accurate ocean forecasting is crucial in different areas ranging from science to decision making. Recent advancements in data-driven models have shown significant promise, particularly in weather forecasting community, but yet no data-driven approaches have matched the accuracy and the scalability of traditional global ocean forecasting systems that rely on physics-driven numerical models and can be very computationally expensive, depending on their spatial resolution or complexity. 
Here, we introduce GLONET, a global ocean neural network-based forecasting system, developed by Mercator Ocean International. GLONET is trained on the global Mercator Ocean physical reanalysis GLORYS12 to integrate physics-based principles through neural operators and networks, which dynamically capture local-global interactions within a unified, scalable framework, ensuring high small-scale accuracy and efficient dynamics. GLONET's performance is assessed and benchmarked against two other forecasting systems: the global Mercator Ocean analysis and forecasting $1/12^{\circ}$ high-resolution physical system GLO12 and a recent neural-based system also trained from GLORYS12.
A series of comprehensive validation metrics is proposed, specifically tailored for neural network-based ocean forecasting systems, which extend beyond traditional point-wise error assessments that can introduce bias towards neural networks optimized primarily to minimize such metrics. The preliminary evaluation of GLONET shows promising results, for temperature, sea surface height, salinity and ocean currents. GLONET's experimental daily forecast are accessible through the European Digital Twin Ocean platform EDITO.
\end{abstract}

\section*{Plain Language Summary}
Accurate ocean forecasting is vital for various applications, from scientific research to decision-making in marine operations. Traditional forecasting systems rely on complex physics-driven numerical models, which can be computationally expensive and slow to improve. GLONET offers a data-driven alternative by utilizing machine learning to predict ocean conditions efficiently. By learning from historical ocean data, GLONET captures intricate ocean dynamics, providing accurate and timely forecasts. Its performance has been validated against existing forecasting systems, showing promising results in predicting key ocean variables.

\newpage
\section{Introduction}
\label{sec:intro}

The accurate prediction of oceanic states is crucial for numerous marine operations, encompassing navigation \cite{james1957application}, fisheries management \cite{pikitch2004ecosystem}, disaster response \cite{teal1984oil, breivik2013advances, zohdi2019harmful}, and climate research. Ocean forecasting has primarily depended on physics-based numerical ocean models \cite{rosati1988general, madec1997ocean, chassignet2007hycom} co-developed and operated by various operational forecasting centers worldwide. These numerical models employ intricate systems of partial differential equations to simulate ocean dynamics, sometimes with the constraint of not departing too far from both contemporary and historical oceanographic observations. Operational Global Ocean Forecasting Systems (GOFSs) merge these numerical frameworks with real-time observational database via advanced data assimilation techniques \cite{bannister2017review, brasseur2005data, cummings2013variational, lellouche2018recent, copernicus2023copernicus} to deliver precise and timely forecasts. However, these models face significant challenges: they are computationally expensive, slow to improve, and require substantial resources for forecasting the state of the ocean, limiting their ability to provide forecasts rapidly, especially in situations that require high-frequency updates or providing ensemble forecasts and their associated uncertainty. This sluggishness also hampers their use in research, where exploring a wide range of scenarios quickly and efficiently can be essential for advancing ocean science. Moreover, even with the influx of more observational data, it is often challenging to improve these models due to their reliance on fixed physical assumptions and the need for extensive parameter tuning. The need for faster, more flexible approaches becomes then evident as traditional models struggle to adapt to evolving requirements and increasingly diverse data sources \cite{liu2017big}.

In recent years, the weather forecasting community has witnessed a significant surge in the development and deployment of neural network (NN)-based forecasting systems. These data-driven models leverage advancements in machine learning, particularly deep learning, to enhance forecast accuracy and computational efficiency. Weather forecasting models such as FourCastNet \cite{pathak2022fourcastnet}, GraphCast  \cite{lam2022graphcast}, Pangu-Weather \cite{bi2023accurate} and recently AIFS \cite{lang2024aifs} have demonstrated remarkable improvements in prediction performance and speed, often outperforming traditional numerical weather prediction (NWP) methods in various metrics \cite{rasp2024weatherbench}. The rapid adoption of NN-based systems in meteorology underscores the potential of artificial intelligence (AI) to revolutionize environmental forecasting.

On the other hand, the oceanographic community faces distinct challenges in developing similar AI-driven forecasting systems. The ocean environment is inherently more complex due to factors like continental boundaries, islands, complex boundary conditions, varying water mass properties, and intricate ocean-atmosphere interactions. These complexities make it significantly more difficult to create data-driven models that can accurately capture the spatial-temporal dynamics of the global ocean. Additionally, the state of AI-based ocean forecasting remains nascent, with limited models achieving the high-resolution forecasts necessary for operational use, typically at $1/4^{\circ}$ or finer, further exacerbates the computational demands, making the training and deployment of deep learning models particularly challenging.

Addressing these challenges, Mercator Ocean International (MOI) has developed a novel neural network-based forecasting system, GLONET, which represents a significant advancement in the field of ocean forecasting. GLONET is trained on the $1/12^{\circ}$ daily mean outputs of the global Mercator Ocean physical reanalysis GLORYS12 \cite{jean2021copernicus} and is designed to operate at a horizontal resolution of $1/4^{\circ}$ while inheriting the fine-scale dynamics of GLORYS12, enabling it to resolve mesoscale oceanic features with unprecedented accuracy. Specifically developed for short-term forecasting up to 10 days, in alignment with operational oceanography protocols, GLONET leverages a hierarchical transformer-based backbone. It incorporates physics-based principles through neural operators and networks to dynamically capture local-global interactions within a unified, scalable framework, mitigating the complexities introduced by continental boundaries and ensuring high small-scale accuracy and efficient dynamics over short forecast intervals.

To ensure the operational viability of GLONET, MOI has established robust pre-operational pipelines that facilitate its integration into existing forecasting workflows. Additionally, GLONET undergoes rigorous validation using the Class-4 framework \cite{ryan2015godae,hernandez2009validation,divakaran2015godae,lellouche2013evaluation}, a stringent evaluation protocol that assesses forecast accuracy against leading operational GOFSs. In addition to traditional metrics, GLONET also undergoes a unique NN-specific evaluation criterion, proposed in this paper, to meet the high standards required for operational deployment. 
In addition, using the aforementioned validation frameworks,  GLONET is benchmarked against the current 
global Mercator Ocean analysis and forecasting $1/12^{\circ}$ high-resolution physical system GLO12 \cite{lellouche2023}, and Xihe, one of the first neural global ocean forecasting system \cite{wang2024xihe} trained from the same ocean reanalysis GLORYS12.

The paper is organized as follows. Section 2 gives an overview of the proposed methodology highlighting its different modules designed to enhance forecast accuracy. Section 3 details the experimental datasets employed in the $1/4^{\circ}$ configuration GLONET along with training. Section 4 presents experimental results, benchmarking GLONET's performance against GLO12 and Xihe, using various validation metrics. Finally, Section 5 concludes the paper with a discussion of the implications of GLONET's performance and potential avenues for future research.

\section{Model}
\label{sec:model}

A diverse range of deep learning architectures such as vision transformers, graph neural networks, and neural operators has found application in data-driven weather forecasting and, more recently, in ocean forecasting \cite{wang2024xihe}, where vision and Swin transformers \cite{khan2022transformers,liu2021swin} were combined to develop a global ocean forecasting system. Building on these advances, we introduce GLONET, a multi-scale forecasting system. The ocean is fundamentally a turbulent system, marked by the interplay of processes across a wide range of spatial and temporal scales. To address this complexity, GLONET leverages Fourier neural operators \cite{li2020fourier} to model broad, basin-wide patterns like gyres, currents, tele-connections, and global circulation, capturing large-scale dependencies and effectively representing the large-scale energy transfer that drives the background state of the ocean. At smaller scales, GLONET employs convolutional neural networks (CNNs) \cite{o2015introduction} to enhance predictions of submesoscale features, which influence localized mixing and fine-scale dependencies. Additionally, GLONET employs an encoder-decoder \cite{schmidhuber2015deep} to fuse these multi-scale circulations into a unified latent space, capturing essential detail across different scales that play a key role in redistributing heat, salt, and energy.

To formalize the learning process, we frame GLONET within an operator learning perspective, which naturally suits the spatio-temporal nature of ocean forecasting. Instead of directly predicting the next ocean state through fixed-size feature transformations, we view forecasting as learning a mapping between function spaces: the space of initial conditions to the space of future states. This formulation allows GLONET to inherently respect the continuous, spatially varying structure of ocean fields, and aligns with the recent trend of modeling physical systems as operators acting on entire fields rather than isolated grid points. This perspective is crucial for modeling the ocean's interconnected behavior, where changes at one location can influence distant regions through wave propagation, advection, and teleconnections.

Let $D \subset \mathbb{R}^d$ define a spatial domain, with input function space $\mathcal{X} = \mathcal{X}(D; \mathbb{R}^{d_X})$  and the output space $\mathcal{Y} = \mathcal{Y}(D; \mathbb{R}^{d_Y})$,  respectively for the initial condition and forecast states, with $\mathbb{R}^{d_Y} \in \mathbb{R}^{d_X}$.

GLONET aimes to approximate the nonlinear mapping $ \mathcal{G}^\dagger : \mathcal{X} \rightarrow \mathcal{Y} $ that forecasts the ocean state $ Y_{t+dt} $ given initial conditions $ X_{t} $.
Let $ \{ (\mathbf{X}_j, \mathbf{Y}_j) \}_{j=1}^N $ be our observed input-forecast pairs, where $ \mathbf{X}_j = X_{t}^j $ represents the initial conditions and $ \mathbf{Y}_j = Y_{t+dt}^j $ the target state. The aim is then to approximate $ \mathcal{G}^\dagger $ using a parameterized operator $ \mathcal{G}_\theta : \mathcal{X} \rightarrow \mathcal{Y} $, where $ \theta \in \Theta $ denotes the parameter set, by minimizing a forecast error $ \mathcal{L} : \mathcal{Y} \times \mathcal{Y} \rightarrow \mathbb{R} $:
\begin{equation}
\min_{\theta \in \Theta} \mathbb{E}_{\mathbf{X} \sim \mu} [\mathcal{L}(\mathcal{G}_\theta(\mathbf{X}), \mathcal{G}^\dagger(\mathbf{X}))]
\end{equation}

where $ \mu $ is a probability measure over the initial conditions function space $ \mathcal{X} $.
In practical terms, this theoretical formulation guides the design of GLONET’s architecture by emphasizing the need for global context, multi-scale feature extraction, and flexible handling of spatially distributed inputs. Rather than treating prediction as a local, pixel-wise task, the operator-based view motivates the explicit modeling of broad spatial dependencies (via Fourier neural operators) alongside fine-grained local dynamics (via CNNs). This dual approach is crucial for improving forecast accuracy in complex systems like the ocean, where processes operate and interact across multiple scales simultaneously.

\begin{figure}[h!]
    \centering
    \includegraphics[width=0.8\textwidth]{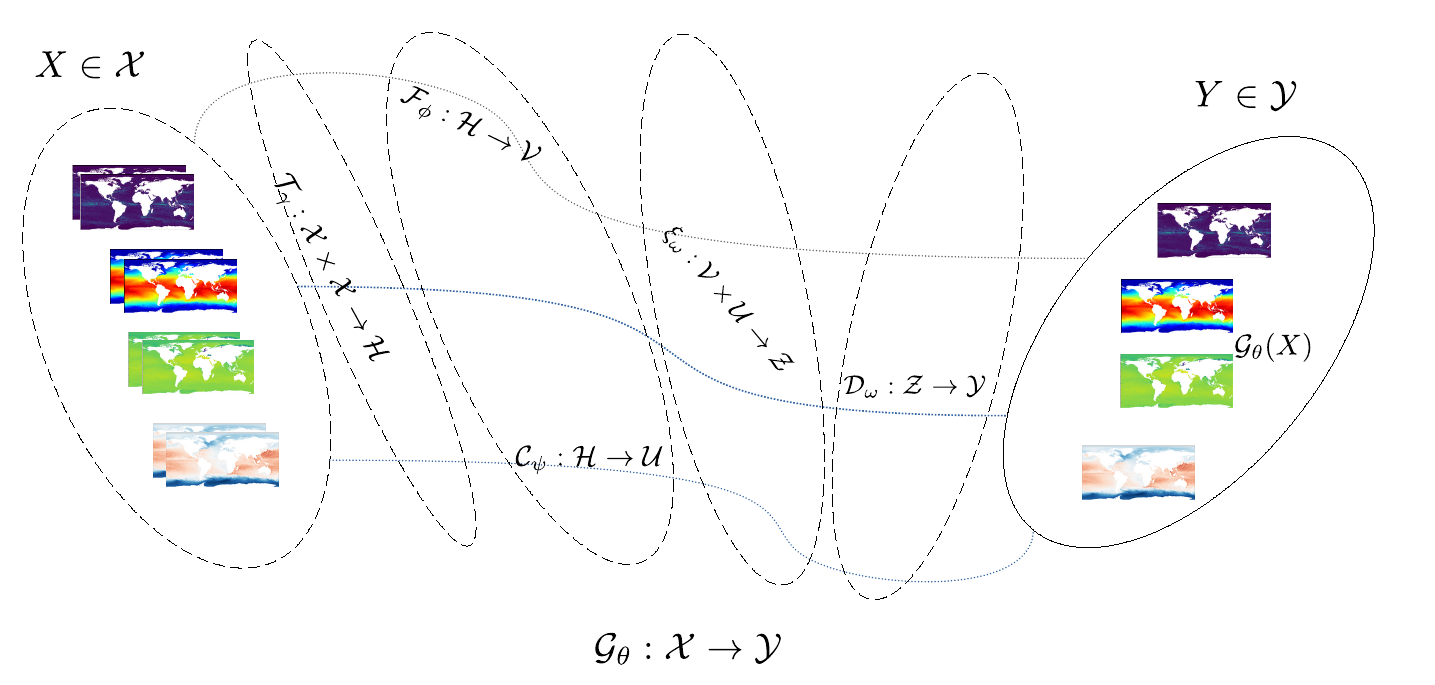}
    \caption{Overview of GLONET's architecture containing different modules, particularly time-block designed to learn feature maps encapsulating initial conditions along with forcings. A spatial module architectured to learn multi-scale dynamics, and finally and encoder-decoder to fuse multi-scale circulations into a unified latent space.\label{fig:glonet}}
\end{figure}

The operator $\mathcal{G}_\theta$ is architected to capture the multi-scale dynamics of ocean forecasting through a series of interconnected components as shown in Figure~\ref{fig:glonet}. 
It starts by processing two consecutive ocean states, $ \mathbf{X}_{t-1} $ and $ \mathbf{X}_{t} $, to extract meaningful feature maps that encapsulate both the initial condition along with the forcings influencing the ocean dynamics as follow:
\begin{equation}
\mathbf{H}_t = \mathcal{T}_\gamma(\mathbf{X}_{t-1}, \mathbf{X}_{t}),    
\end{equation}

where:
\begin{itemize}
    \item $ \mathcal{T}_\gamma : \mathcal{X} \times \mathcal{X} \rightarrow \mathcal{H} $ is the temporal encoding operator parameterized by $ \gamma $.
    \item $ \mathcal{H} $ represents the feature space that includes both the current ocean state and the learned forcings.
\end{itemize}
This block effectively integrates temporal information and forcings, providing a rich representation for subsequent spatial processing. To capture the diverse spatial scales inherent in the ocean, $\mathcal{G}_\theta$ employs processing through two distinct modules $ \mathcal{F}_\phi $  and $ \mathcal{C}_\psi $. 
The first is designed to model large-scale, basin-wide ocean patterns by capturing global dependencies and long-range interactions,  providing a foundational representation of the ocean state. It operates in the Fourier domain, leveraging the Fourier transform $ \mathcal{F} $ and its inverse $ \mathcal{F}^{-1} $ to efficiently capture spectral information as:
\begin{equation}
\mathbf{v}_{t+1} = \mathcal{F}_\phi(\mathbf{H}_t) = \mathcal{F}^{-1} \left( \sigma \left( W \cdot \mathcal{F}(\mathbf{H}_t) \right) \right),
\end{equation}

with $ W \in \mathbb{C}^{k_{\text{max}} \times d_H \times d_H} $ are learned weights in the Fourier space,  $ \sigma $ is a non-linear activation function applied element-wise, and $ k_{\text{max}} $ defines the maximum frequency mode considered. 
Simultaneously, $ \mathcal{C}_\psi $ focuses on capturing small-scale, localized features and interactions. It operates directly in the spatial domain, applying convolutional filters to extract fine-grained patterns and defined as:
\begin{equation}
\mathbf{u}_{t+1} = \mathcal{C}_\psi(\mathbf{Y}_t),
\end{equation}

where $ \mathcal{C}_\psi : \mathcal{F} \rightarrow \mathcal{U} $ is a convolutional neural networks parameterized by $ \psi $, and $ \mathcal{U} $ the feature space capturing localized dependencies. 
The combination of Fourier-based global modeling and CNN-based local refinement directly addresses the dual nature of ocean dynamics: the need to accurately forecast both broad circulation patterns and localized phenomena such as eddies and fronts. Without the operator-based view, such explicit separation and integration of scales would be much harder to achieve within a single model.

To integrate the large-scale and small-scale feature representations, $\mathcal{G}_\theta$ employs an encoder-decoder that fuses multi-scale circulations from $ \mathbf{v}_{t+1} $ and $ \mathbf{u}_{t+1} $ into a unified latent representation:

\begin{equation}
\mathbf{z}_{t+1} = \mathcal{E}_\omega(\mathbf{v}_{t+1}, \mathbf{u}_{t+1}),
\end{equation}

where:
\begin{itemize}
    \item $ \mathcal{E}_\omega : V \times U \rightarrow Z $ is the encoder operator parameterized by $ \omega $.
    \item $ Z $ represents the latent feature space that encapsulates both large-scale and small-scale dynamics.
\end{itemize}

On the other hand, $ \mathcal{D}_\omega $ maps the latent representation $ \mathbf{z}_{t+1} $ back to the original spatial domain, generating the forecasted ocean state $ \mathbf{X}_{t+dt} $:

\begin{equation}
\mathbf{X}_{t+dt} = \mathcal{D}_\omega(\mathbf{z}_{t+1}),
\end{equation}

with $ \mathcal{D}_\omega : Z \rightarrow \mathcal{Y} $ is a decoder operator parameterized by $ \omega $.

Combining all components, $\mathcal{G}_\theta$ can be expressed as the composition of all the aforementioned operators:
\begin{equation}
\mathcal{G}_{\theta}(\mathbf{X}_{t-1}, \mathbf{X}_t) = \mathcal{D}_\omega \left( \mathcal{E}_\omega \left( \mathcal{F}_\phi \left( \mathcal{T}_\gamma(\mathbf{X}_{t-1}, \mathbf{X}_t) \right), \mathcal{C}_\psi \left( \mathcal{T}_\gamma(\mathbf{X}_{t-1}, \mathbf{X}_t) \right) \right) \right),
\end{equation}

where $ \theta = \{\gamma, \phi, \psi, \omega\} $ encapsulates all model parameters across the different modules. This composite mapping effectively captures the multi-scale dynamics of the ocean by integrating temporal dependencies, large-scale circulations, and localized features into a unified forecasting framework.

To efficiently handle the high-resolution requirements of GLONET, a multi-GPU pipeline that leverages pipeline parallelism is adopted, in which distinct segments of the model are distributed across different GPUs, passing activations between devices as the model progresses through the pipeline. This approach ensures that the computational load is shared and enables scaling to larger model sizes while accommodating high-resolution data. Additionally, activation checkpointing is employed throughout the forward pass to further reduce memory usage. Instead of keeping intermediate activations, which can quickly consume GPU memory, intermediate results are selectively recomputed during the backward pass. This strategy significantly minimizes the memory footprint and reduces the peak memory requirement, allowing each GPU to handle larger parts of the model without exceeding memory limitations. 
Furthermore, to facilitate the above, GPU operations are synchronized using efficient communication primitives, enabling a smooth transfer of activations between GPUs with minimal overhead. By distributing both the model parameters and activations across multiple devices, GLONET is trained on fine-resolution datasets without compromising computational efficiency.

\section{Data and Training}
\label{sec:training}

\subsection{Data}
GLORYS12 reanalysis \cite{jean2021copernicus} provides a gap-free, physically consistent reconstruction of the global ocean state, developed by Mercator Ocean International as part of the Copernicus Marine Service. GLORYS12 combines numerical ocean models with in situ and satellite observations through a sophisticated data assimilation scheme, ensuring that the reanalysis remains faithful to observations while maintaining dynamical consistency. It delivers daily, three-dimensional fields of temperature, salinity, currents, and sea surface height at a horizontal resolution of $1/12^\circ$, supporting the needs of operational oceanography. This high-quality dataset serves as a foundation for the training, validation, and evaluation of GLONET.

\subsection{Training}
GLONET is trained on daily mean outputs from the GLORYS12 reanalysis and is configured to produce daily forecasts using the ocean state variables listed in Table \ref{tab:variables}. The selected vertical levels are designed to sample key dynamical regimes across the full ocean column. The first four levels, located at approximately 0.49 m, 47 m, 92 m, and 155 m, reside within the upper 200 meters and capture near-surface variability, the seasonal mixed layer, and the upper thermocline, regions central to air-sea interactions, mixed layer dynamics, and stratification onset. The intermediate levels (222 m to 541 m) sample the deeper thermocline and upper intermediate waters, where vertical gradients remain pronounced and contribute to processes such as subsurface ventilation and water mass transformation. The subsequent range from 643 m to 1245 m spans the lower thermocline into the upper deep ocean, where vertical variability decreases but key features, including the base of the thermocline and intermediate water masses, continue to exert dynamical influence. Levels deeper than 1245 m, extending to 5274 m, are included to represent the structure of the abyssal ocean and to account for large-scale circulation patterns, stratification, and bathymetric effects.

While the direct impact of deep ocean conditions on upper-ocean forecasts at a 10-day timescale is expected to be limited, the inclusion of full-depth profiles during training allows the model to learn cross-depth statistical relationships and to better capture low-frequency variability such as seasonal cycles and basin-scale dynamics. Additionally, by exposing the model to deep ocean structure, it is indirectly informed about the influence of bathymetry and topographic constraints on the general circulation, particularly in regions where bottom features steer currents or shape mesoscale variability. This vertically comprehensive configuration supports better generalization and enhances the physical consistency of the forecasts within a unified, data-driven framework.

To optimize computational efficiency, GLONET operates on a coarser $[1/4]^\circ$ grid, by interpolating GLORYS12 products to this resolution. This decision is motivated by the substantial increase in memory requirements associated with high-resolution spatiotemporal models, where the memory footprint grows approximately quadratically with the spatial resolution. Moreover, during training, additional memory overhead arises from the need to store intermediate activations for backpropagation and gradient computation, further exacerbating resource demands. Interpolating the inputs to $[1/4]^\circ$ thus enables efficient training and inference while preserving the key mesoscale dynamics intrinsic to the original $[1/12]^\circ$ products, ensuring a balance between computational tractability and spatial fidelity.

Importantly, in physics-based numerical models, the grid resolution imposes a fundamental constraint on the smallest scales that can be dynamically resolved, with coarse resolutions inherently limiting the representation of mesoscale and submesoscale processes. In contrast, machine learning models trained on interpolated high-resolution data are not governed by the same dynamical restrictions. In the present configuration, interpolating the $[1/12]^\circ$ GLORYS12 fields to a $[1/4]^\circ$ grid preserves a substantial portion of the small-scale variability, as confirmed through both spectral analyses and qualitative assessments. This stands in contrast to physical models natively run at $[1/4]^\circ$ resolution, where the absence of finer grid spacing results in a significant smoothing of the fields and a degradation of mesoscale features. Consequently, the interpolation strategy employed in GLONET enables the model to remain exposed to rich mesoscale dynamics during training, despite operating at a coarser computational resolution, thereby maintaining a critical balance between computational efficiency and physical realism.

\begin{table}[h]
    \centering
    \begin{tabular}{lcc}
        \toprule
        \textbf{Variable} & \textbf{Depth Levels [m]} & \textbf{Spatial Resolution} \\
        \midrule
        Temperature & 0.49, 47, 92, 155 & $[1/4]^{\circ}$ (input/output) \\
        Salinity & 222, 318, 380, 453, 541 & $[1/4]^{\circ}$ (input/output) \\
        U and V current components & 643, 763, 902, 1245, 1684, & $[1/4]^{\circ}$ (input/output) \\
        & 2225, 3220, 3597, 3992, 4405, 4833, 5274 & \\
        SSH & Surface & $[1/4]^{\circ}$ (input/output) \\
        \bottomrule
    \end{tabular}
    \caption{GLONET configuration: input/output variables, their corresponding depth levels, and spatial resolution. U and V correspond to the zonal and meridional velocity components, respectively. SSH corresponds to the sea surface height.}
    \label{tab:variables}
\end{table}

GLONET forecasts the three-dimensional ocean state at  $t_{+1day}$  by leveraging the ocean states at $t$ and $t_{-1day}$, thanks to its temporal module $\mathcal{T}_{\gamma}$, which integrates informative feature maps that encapsulate both the initial conditions and the external forcings modulating ocean dynamics.
GLONET $\mathcal{G}_{\theta}(\mathbf{X}_{t-1}, \mathbf{X}_t)=\mathbf{X}_{t+1}$ is designed to provide 1 lead-day forecast. Therefore, longer lead-time forecasts are produced through an auto-regressive approach, where the GLONET iteratively inputs its own predictions. To enhance forecast accuracy, the same process is also considered during the training phase, where an initial pre-training phase focuses on generating 1-day-ahead forecasts, followed by a subsequent phase in which the GLONET extends forecasts up to 4 days. During backpropagation, gradients propagate across the entire 4-days forecast sequence, ensuring that temporal dependencies are captured and optimized throughout the training process.

The training set consists of daily mean GLORYS12 reanalysis outputs from 1993 to 2019. The year 2020 is reserved for validation, providing an unseen dataset for tuning hyperparameters and monitoring generalization during training. For the final evaluation, GLONET is tested using independent initial conditions derived from the operational forecast products GLO12 to rigorously assess out-of-sample performance. During training, we employ random shuffling of samples to enhance statistical robustness and prevent potential temporal or spatial biases within each batch, ensuring more efficient gradient updates and faster convergence.

The training of GLONET involves optimizing the model parameters $ \theta $ to minimize the expected forecast error over the distribution $ \mu $ of initial conditions, derived from GLORYS12 reanalysis. The optimization problem is formulated as:
 \begin{equation}
\min_{\theta \in \Theta} \mathbb{E}_{\mathbf{X} \sim \mu} \left[ \mathcal{L} \left( \mathcal{G}_\theta(\mathbf{X}), \mathcal{G}^\dagger(\mathbf{X}) \right) \right],
 \end{equation}

where $ \mathcal{L} $ is a suitable loss function (e.g., mean squared error) that quantifies the discrepancy between the predicted state $ \mathcal{G}_\theta(\mathbf{X}) $ and the true state $ \mathcal{G}^\dagger(\mathbf{X}) $. In practice, this expectation is approximated using an empirical average over a finite training dataset, sampled according to $ \mu $ and based on GLORYS12 daily mean states:

\begin{equation}
\min_{\theta \in \Theta} \frac{1}{N} \sum_{j=1}^N \mathcal{L} \left( \mathcal{G}_\theta(\mathbf{X}_j), \mathbf{Y}_j \right).
\end{equation}

This empirical risk minimization enables GLONET to learn a robust and generalizable forecasting model, accurately predicting the ocean state under varying initial conditions derived from high-resolution reanalysis data. The GLORYS12 outputs provide an extensive and consistent dataset, facilitating the model's ability to capture the essential small and large-scale ocean dynamics needed for reliable forecasts.

Training was performed on $1/4^{\circ}$ interpolated GLORYS12 daily mean outputs for the period $[1993, 2019]$, using Adam's optimizer \cite{kingma2014adam}, and a learning rate decreasing from $e^{-4}$ to $e^{-5}$. Once the model is trained and the desired performances were reached, a second round of fine-tuning is performed to optimize the forecasts accurary up to 4 days. The overall training took around 3 weeks while using a total of 32 40G-A100 GPUs. GLONET is capable of outputting 10 days forecast in less than 10 seconds.

\section{Results}
\label{sec:results}

In this section, GLONET's forecasting capabilities are empirically assessed using a comprehensive set of validation metrics, designed to capture different dimensions of forecast quality. The evaluation begins with the IV-TT Class 4 Framework, a robust benchmark that leverages observational data to evaluate forecasting accuracy against reference datasets. This is followed by point-wise accuracy assessments, process-oriented metrics, and diagnostic analyses, offering a multifaceted understanding of the model's performance.

The models used in this evaluation differ significantly in their design and application. GLONET, a data-driven model, employs an autoregressive structure, generating forecasts based on prior predictions. In contrast, Xihe, also a data-driven approach, does not rely on autoregressive modeling and instead focuses on independent, single-step forecasts, which provides stable performance across lead times but may lack long-term consistency. GLO12, a physics-driven model, integrates governing physical equations and observational data assimilation, allowing it to perform well in dynamically consistent predictions. Each model differs in resolution and computational efficiency, with GLONET and Xihe benefiting from a more flexible, data-driven architectures that can be trained to resolve high-dimensional inputs, while GLO12 rely on more traditional, computationally intensive physics-based and autoregressive fashion.

\begin{table}[h!] 
\centering 
\begin{tabular}{|l|l|l|l|l|}
\hline 
\textbf{Model} & \textbf{Type} & \textbf{Autoregressive} & \textbf{Computational Efficiency} & \textbf{Resolution} \\
\hline
GLONET & Data-driven & Yes & few seconds (GPU) & $1/4^{\circ}$ \\
\hline 
Xihe & Data-driven & No & few seconds (GPU) & $1/12^{\circ}$  \\
\hline
GLO12 & Physics-based & Yes & 1h (HPC) & $1/12^{\circ}$  \\
\hline
\end{tabular} 
\caption{Comparison of model types, autoregressive features, computational efficiency, and resolution.}
\end{table}

Point-wise evaluations use GLORYS12 as the reference dataset, assessing the models’ ability to generate forecasts that align closely with the data on which they were trained. This analysis provides a detailed view of the models' accuracy in capturing ocean states across various spatial and temporal scales, ensuring consistency and comparability with the training reference.
Extending beyond these localized evaluations, process-oriented metrics are incorporated to evaluate the model's ability to derive essential ocean quantities and maintain coherence across interconnected forecasted variables. Additionally, diagnostic variables, such as vorticity and mesoscale eddies are analysed to assess the model’s skill in preserving dynamic consistency, energy cascading and fine-scale structures. This holistic evaluation ensures that GLONET’s forecasts not only minimize numerical errors but also adhere to the fundamental dynamics governing ocean systems.

\subsection{IV-TT Class 4 Framework Evaluation}

In this part, GLONET, Xihe and GLO12 are evaluated using the CLASS4 dataset, developed and maintained by the "OceanPredict Task Team for Intercomparison and Validation" (IV-TT team). 
The CLASS4 framework provides a benchmark for model validation by operating within the observation space, enabling a direct comparison between observed and modeled values across both spatial and temporal dimensions. For each observation, the corresponding model counterpart is extracted at the same spatial and temporal location across various forecast lead times, ranging from the best analysis (day 0) to ten-day forecasts. 

The CLASS4 dataset includes observations of temperature and salinity from Argo profiles, sea surface temperature (SST) from surface drifting buoys, sea level anomaly (SLA) from along-track satellite measurements and surface current observations at 15m depth from the Global Drifter Program (GDP) drifters buoys. This framework serves as a robust tool for intercomparison, facilitating a comprehensive assessment of the forecasting models' performance. For SST, it is important to note that comparisons are made between the observed SST (from surface drifting buoys) and the modeled SST at the first vertical level (which is at 0.49m depth) for each model. This level is the closest available approximation of the sea surface in the model outputs.

The evaluation of model performance is based on two common metrics: the Root Mean Square Difference (RMSD) and the Mean Absolute Error (MAE).

The equation for RMSE is given by:

\begin{equation}
\text{RMSE} = \sqrt{\frac{1}{N} \sum_{i=1}^{N} (y_i - \hat{y}_i)^2}
\end{equation}

where $ y_i$  represents the observed value, $\hat{y}_i$  the modeled value, and $N$  is the total number of data points.

Similarly, the MAE is calculated as:

\begin{equation}
\text{MAE} = \frac{1}{N} \sum_{i=1}^{N} |y_i - \hat{y}_i|
\end{equation}

where $|y_i - \hat{y}_i|$ is the absolute difference between the observed and modeled values.

\begin{figure}[h!]
    \centering
    \includegraphics[width=1\textwidth]{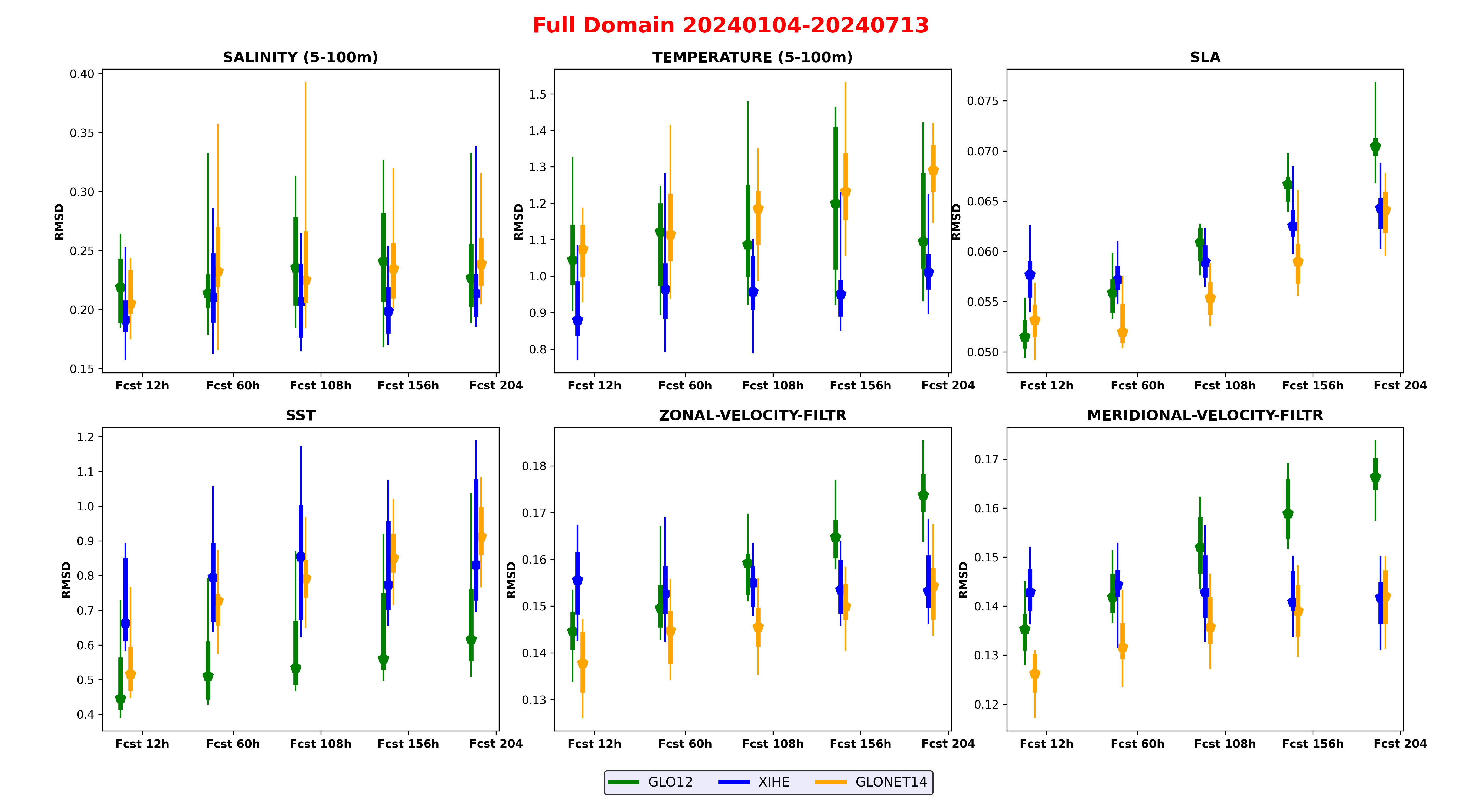}
    \caption{Dispersion and evolution of RMSD as a function of forecast lead time for salinity (top left) and temperature (top middle) in the 5–100 m layer, SLA (top right), and SST/(Temperature of the first level for models) (bottom left) from drifting buoys and zonal and meridional ocean currents (middle of bottom right). The thick lines represent the $75\%$ distribution, while the thin lines correspond to the $95\%$ distribution, the dot represent the median of the distribution \label{fig:class4days}}
\end{figure}

Figure~\ref{fig:class4days} illustrates the dispersion and evolution of the Root Mean Square Difference (RMSD) as a function of forecast lead time for the following variables: temperature, salinity (5–100 m layer), SLA, SST, zonal and meridional surface currents. XIHE performs comparably to GLO12 and slightly better than GLONET in forecasting temperature of the 5–100 m layer. The three models exhibit similar performance in salinity, though GLONET shows greater dispersion for the 3 day and 5 day forecasts. GLONET surpasses XIHE in SLA and surface currents, consistently outperforming GLO12 for forecasts ranging from 5 to 9 days. For SST, GLO12 demonstrates superior performance across all forecast horizons, except for the one-day forecast, where the three models yield identical results. For surface currents, GLONET achieves the highest performance in both meridional and zonal components, indicating that global ocean circulation is well captured in its surface layers relative to GDP observations at 15 m depth.

\begin{figure}[h!]
    \centering
    \includegraphics[width=1\textwidth]{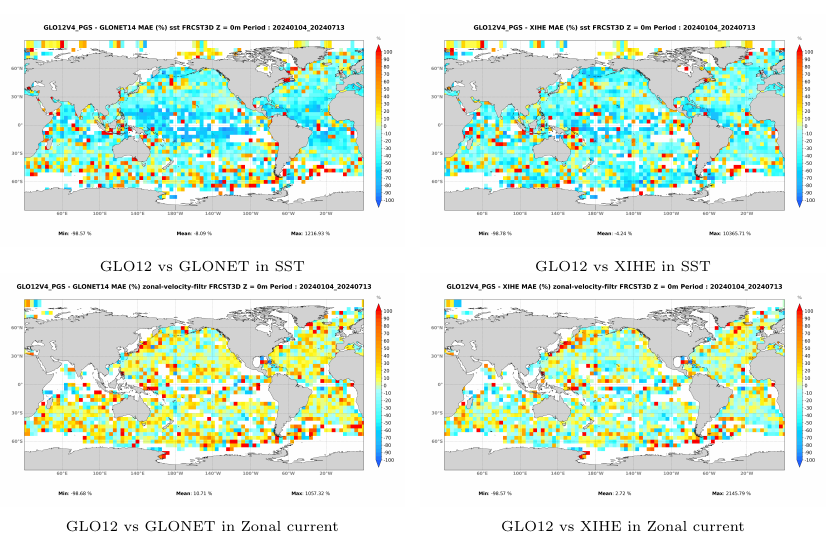}
    \caption{Normalized MAE difference for SST/(Temperature of the first level for models) and zonal current between GLO12 and GLONET (left panels) and XIHE (righ panels). Positive values (red) denote improvement relative to GLO12, while negative values (blue) indicate degradation. \label{fig:class4maps}}
\end{figure}

To further explore regional performance differences, Figure~\ref{fig:class4maps} presents the spatial distribution of normalized Mean Absolute Error (MAE) differences. Here, the MAE is first computed for GLO12 as a reference, followed by the MAE calculations for GLONET and XIHE. Each map shows the difference between the reference and the model, normalized by the reference MAE. Positive values (red) indicate areas where the model outperforms GLO12, while negative values (blue) represent regions of degradation relative to the reference.

For SST the neural network-based models, GLONET and XIHE, generally perform well in dynamically active regions such as the Kuroshio, Gulf Stream, Antarctic Circumpolar Current (ACC), and parts of the North Atlantic and North Pacific. However, both models show reduced performance in the tropical band and within subtropical gyres. GLONET outperforms XIHE in the South Pacific, potentially due to better generalization in regions with sparse observations or lower variability in training data. Conversely, XIHE demonstrates slightly better performance in the North Atlantic, likely benefiting from denser observational data in this well-monitored region. 

For surface currents, GLONET demonstrates robust performance across most regions, with mild localized performance reductions observed in the North and South Pacific Gyre and equatorial zones. GLONET exhibits superior accuracy compared to XIHE in all major ocean basins, including the Pacific, Atlantic, and Indian Oceans. Notably achieving exceptional results in dynamically active regions, such as the Antarctic Circumpolar Current (ACC), where it captures strong current features with high fidelity.

When compared to GLO12, the neural network models reveal nuanced strengths and weaknesses. GLONET and XIHE may struggle in regions dominated by strong seasonal or sub-surface dynamics, as these are more explicitly represented in physics-based models like GLO12. However, in regions characterized by mesoscale variability, such as the Kuroshio and Gulf Stream, the data-driven approaches can leverage their flexibility to capture localized patterns, occasionally outperforming GLO12. These results highlight the complementary strengths of neural network-based and physics-based approaches in forecasting regional ocean dynamics.

\subsection{Point-wise evaluation}

In this part, the point-wise forecasting accuracy of GLONET is evaluated by comparing its outputs against GLORYS12 reanalysis products, which serve as the reference. GLORYS12 provides high-quality, observationally informed ocean state estimates, making it an established benchmark for assessing forecast performance in ocean modeling.

To contextualize GLONET's forecasting capabilities within the domain of data-driven modeling, its 10-day forecasts are compared with those of Xihe, another neural network-based forecasting system. Both GLONET and Xihe are evaluated using identical initial conditions, the same GLORYS12 reference fields, and over a consistent evaluation period spanning January to July 2024. This setup ensures a fair and uniform comparison, enabling an objective assessment of GLONET’s performance relative to other data-driven approaches.

The evaluation of point-wise accuracy is based on the Root Mean Square Error (RMSE), a standard metric for quantifying the difference between observed and predicted values. The RMSE is calculated for each forecast lead time and for each of the model variables (temperature, salinity, and surface current components). For each model and each variable, the RMSE is computed as:

\begin{equation}
\text{RMSE} = \sqrt{\frac{1}{N} \sum_{i=1}^{N} (y_i - \hat{y}_i)^2}
\end{equation}

where $y_i$ are the observed values from the GLORYS12 dataset, $\hat{y}_i$ are the forecasted values from the model (GLONET or Xihe), and $N$ is the number of data points over the evaluation period.

This RMSE-based evaluation allows for a precise comparison of the forecasting accuracy of GLONET and Xihe, highlighting differences in their ability to replicate observed ocean states across various temporal scales and forecast lead times.

\begin{figure}[h!]
    \centering
    \includegraphics[width=1\textwidth]{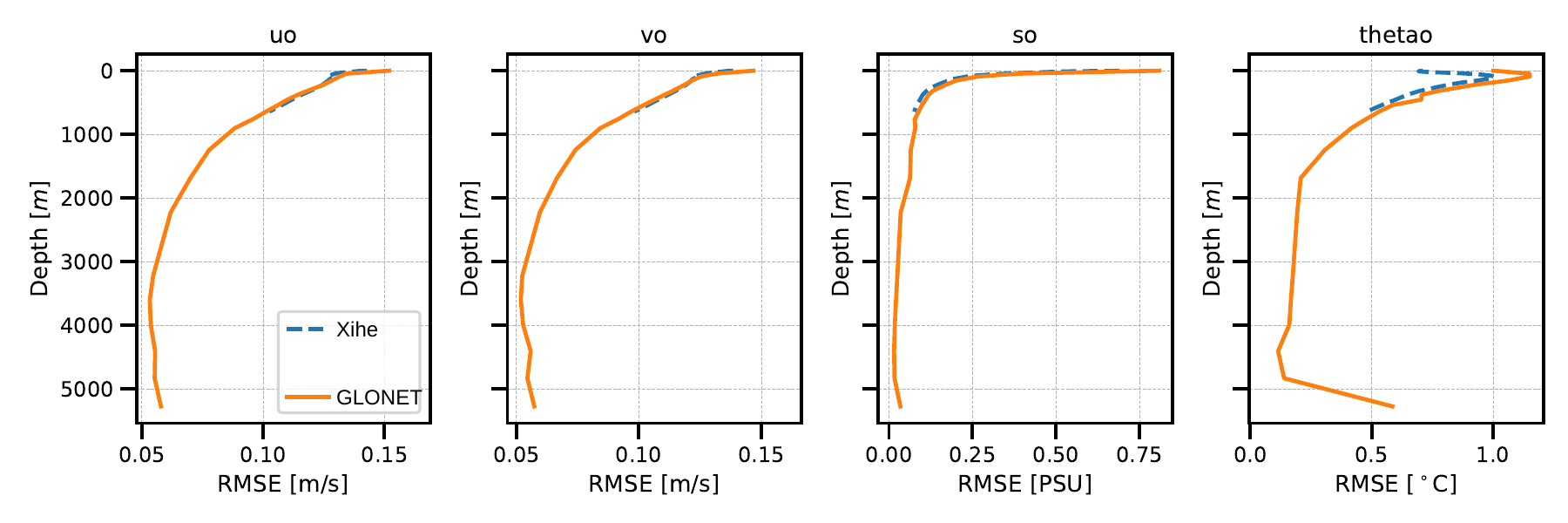}
    \caption{RMSE computed at each depth level and averaged over all lead times for 3D variables (U and V currents, temperature, and salinity) for the GLONET (orange line) and Xihe (blue line) models, covering January to July 2024. GLORYS12 serves as the reference, with 10-day forecasts initialized weekly on Wednesdays from a nowcast analysis performed with GLO12 seven days behind real-time.\label{fig:depthrmse}}
\end{figure}

Figure \ref{fig:depthrmse} shows the RMSE scores averaged over all lead times at each vertical level, highlighting variations in forecasting accuracy with depth. In terms of ocean currents, GLONET consistently outperforms Xihe across most depths, while at surface and mixed-layer depths, the performances are comparable. For temperature, GLONET and Xihe exhibit similar accuracies in the intermediate ocean layers, with slight differences at the surface and bottom levels. This depth-resolved evaluation provides a more detailed understanding of each model’s strengths and weaknesses across the vertical profile.

\begin{figure}[h!]
    \centering
    \includegraphics[width=1\textwidth]{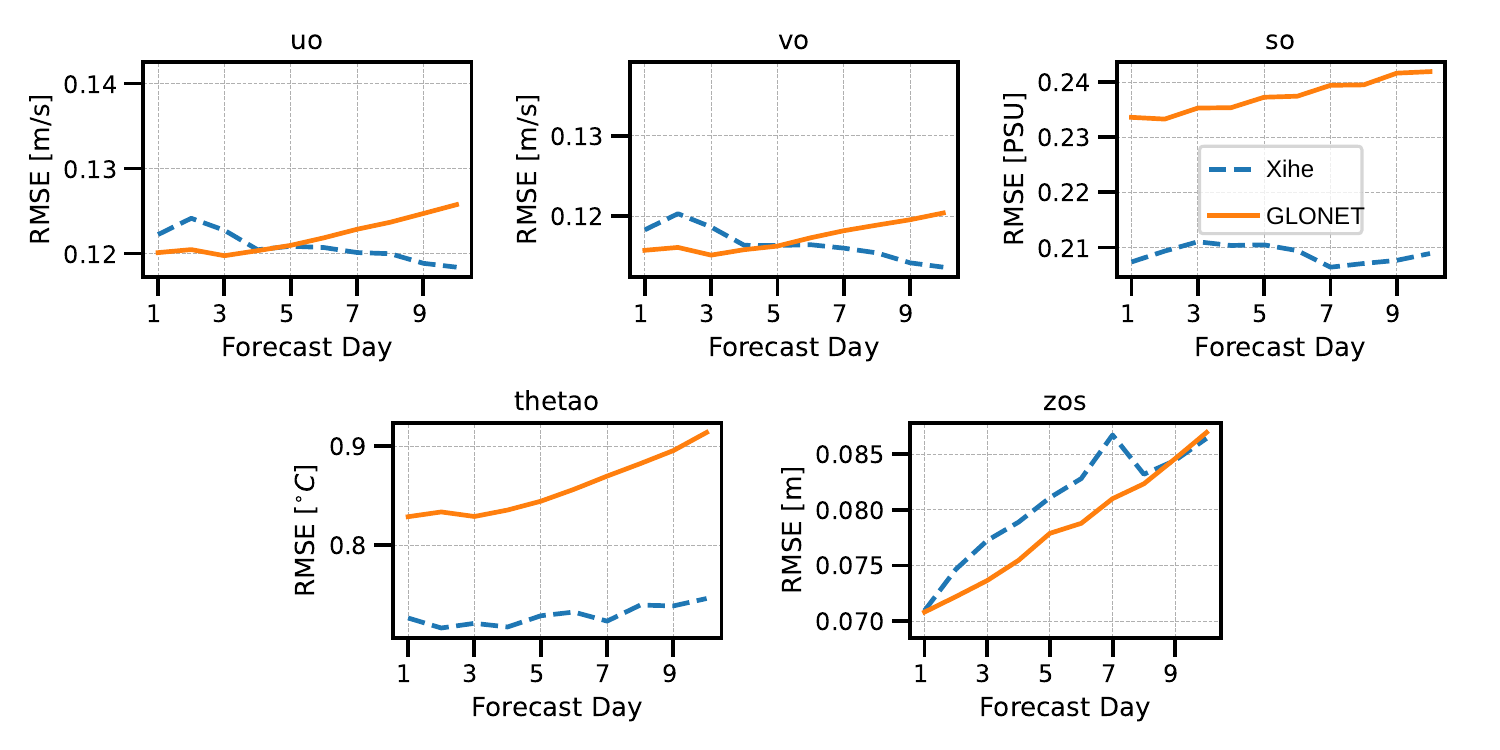}
    \caption{RMSE averaged across common depth levels of 3D variables (U and V currents, temperature, salinity), along with RMSE of sea surface height (SSH) for the GLONET and Xihe models. Calculations span from January to July 2024, using GLORYS12 as the reference. Forecasts are initialized weekly on Wednesdays, following the operational protocol.\label{fig:dayrmsed}}
\end{figure}

To ensure a fairer comparison given differences in vertical discretizations between models, RMSE scores are also averaged across the common depth levels shared by GLONET and Xihe, as shown in Figure \ref{fig:dayrmsed}. This analysis confirms that GLONET maintains strong performance for ocean currents throughout the forecast window. In terms of temperature, GLONET and Xihe perform similarly at intermediate depths, while Xihe shows a slight advantage near the surface and at the bottom layers.

\begin{figure}[!htbp]
    \centering
    \includegraphics[width=\textwidth]{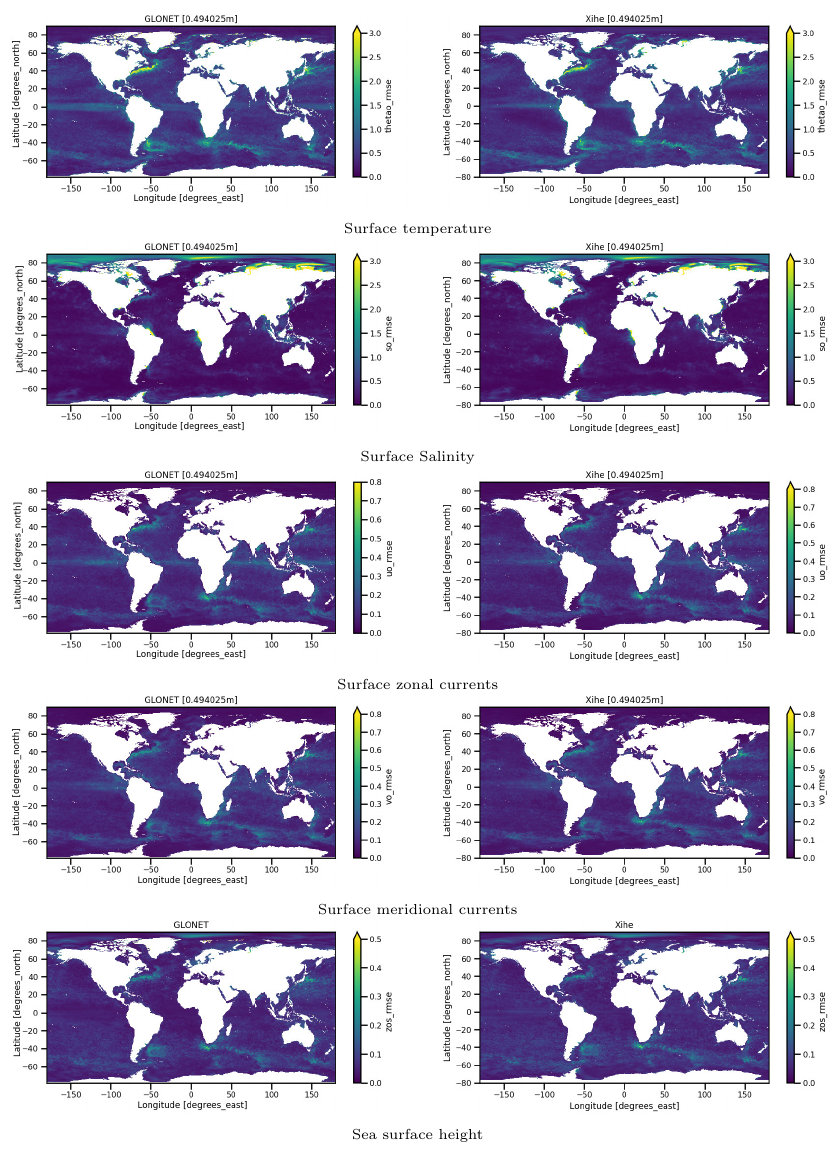}
    \caption{Spatial distribution of RMSE at the surface level for GLONET (left) and Xihe (right) across all forecasted variables (U and V currents, temperature, salinity, and SSH). The RMSE is computed with respect to GLORYS12 over the evaluation period from January to July 2024. Results are averaged over all lead times for weekly forecasts initialized on Wednesdays, adhering to the operational protocol.}
    \label{fig:maprmse}
\end{figure}

In addition to the analyses presented earlier, a complementary evaluation focusing on the spatial distribution of forecasting errors at the surface level is provided. This analysis aims to assess the regional variability of model performance, offering insights into how well GLONET and Xihe capture the spatial patterns of ocean dynamics. The spatial RMSE maps at the surface level, shown in Figure \ref{fig:maprmse}, reveal that both models, GLONET and Xihe, exhibit a broadly similar regional distribution of errors across all forecasted variables (U and V currents, temperature, and salinity). This consistency in spatial patterns suggests that the primary sources of forecasting error are shared between the models, likely driven by the underlying ocean dynamics and the challenges associated with accurately resolving certain high-variability regions, such as western boundary currents and equatorial zones. While the magnitude of RMSE varies, the spatial consistency underscores the difficulty of forecasting in these regions regardless of the modeling approach.

\subsection{Process-oriented evaluations}
In this part, GLONET is evaluated using a process-oriented approach, which involves computing derived quantities and analyzing complex processes from the forecasted data. Unlike physics-driven models, which rely on governing equations to represent ocean dynamics, data-driven models are primarily trained to minimize a loss function, typically based on RMSE. Consequently, neural network-based models are optimized to achieve the best point-wise accuracy but are not inherently guided by the physical laws that govern ocean system. 
This limitation necessitates additional forms of validation to ensure that the models not only achieve low error metrics but also respect the underlying dynamics and interactions within the ocean system. By deriving critical oceanographic quantities such as mixed layer depth and geostrophic currents from the models forecasted data, the models abilities to capture realistic physical relationships within the ocean is assessed. Moreover, this process-oriented evaluation provides insight into the internal consistency and coherence across forecasted variables, verifying that the interactions among temperature, salinity, currents, and sea surface height are physically plausible and aligned.
Such coherence is crucial for oceanographic forecasting, as the interplay between these variables underpins accurate representation of the complex ocean processes. A lack of alignment could signal inconsistencies that arise from the data-driven models which focus on minimizing RMSE rather than capturing these interdependent dynamics. This dual assessment of derived quantities and variable consistency thus provides a rigorous evaluation of neural-based forecasting systems capacity to not only deliver accurate forecasts on an individual variable basis but also to maintain the integrity of the broader ocean system, offering a holistic view of the models abilities to replicate realistic ocean dynamics.

\subsubsection{Derived quantities}
In this part, two essential physical properties are derived: geostrophic currents \cite{rhines1979geostrophic} and mixed layer depth (MLD) \cite{de2004mixed} from the forecasted variables to assess the physical consistency, dynamic accuracy, and potential artifacts in the models' outputs. The evaluation of these derived quantities is reported in terms of the Root Mean Square Error (RMSE) to quantify the discrepancy between the reference and model-predicted values. This allows for an objective comparison of the models' ability to replicate these key physical features, ensuring their dynamical accuracy and overall consistency with observed ocean behaviors.

 \paragraph{Geostrophic currents} provide a diagnostic of large-scale ocean circulation. Accurate prediction of these currents is critical for understanding ocean transport and dynamics. They are derived from forecasted SSH under the geostrophic approximation:
 \begin{equation}
     \mathbf{v}(\phi, \theta, t) =gf^{-1}\nabla^{\perp}\eta(\phi, \lambda, t)
 \end{equation}
where $g$ is the acceleration of gravity, $f$ presents the Coriolis coefficient, and $\eta(\phi, \lambda, t)$ is the sea surface height (SSH), which serves as a noncanonical Hamiltonian for surface velocity. $\perp$ stands for a $90^{\circ}$ anticlockwise rotation of the gradient vector, producing a perpendicular flow direction as dictated by geostrophic balance.

\begin{figure}[h!]
    \centering
    \includegraphics[width=1\textwidth]{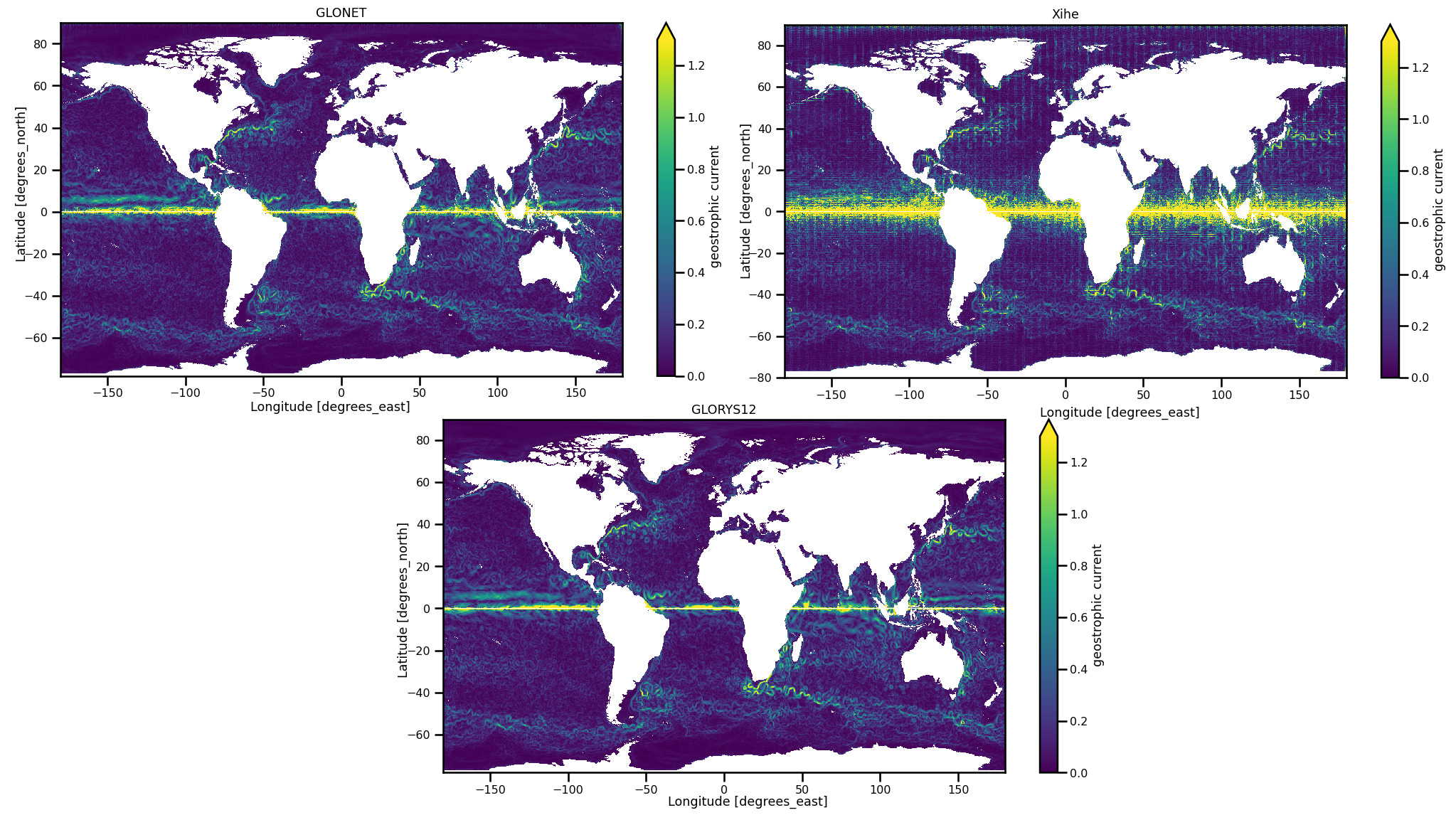}
    \caption{ Examples of approximated geostrophic currents using outputs of GLONET and Xihe over the same date (05-01-2024) along with GLORYS12 serving as a reference. \label{fig:geo}}
\end{figure}

Figure ~\ref{fig:geo} presents examples of derived geostrophic currents computed from GLONET, Xihe, and GLORYS12, the latter serving as a reference. GLONET's derived geostrophic currents show very clean results, reproducing structures and patterns nearly identical to those in GLORYS12. This indicates a high degree of physical consistency in GLONET's SSH forecasts, capturing the expected gradient flows with minimal distortion. The similarity in structure between GLONET and GLORYS12 further attests to GLONET's capability in maintaining dynamic accuracy even when evaluated through derived, process-oriented quantities. In contrast, Xihe's derived geostrophic currents appear notably noisy, with visible artifacts in the output. These artifacts are largely concealed in direct SSH predictions but become evident when calculating the geostrophic current, a process that involves spatial gradients. Since gradient-based calculations can amplify inconsistencies or abrupt changes, Xihe's noise suggests underlying instabilities or artifacts in the SSH field that would otherwise remain undetected. 

\paragraph{Mixed Layer Depth} (MLD) is derived from forecasted temperature and salinity profiles and serves as a key indicator of upper-ocean stratification and mixing. Accurately predicting MLD is essential for simulating air-sea interactions, heat exchange, and biological productivity. MLD is commonly defined based on a density threshold criterion, such that the mixed layer is the depth at which the density difference from the surface equals a specified threshold. The MLD can be approximated as:
\begin{equation}
    \text{MLD} = \min \left\{ z \mid \rho_z - \rho_0 \geq \Delta \rho \right\}
\end{equation}
where $ \rho_z $ represents the density at depth $ z $, $ \rho_0 $ is the density at the surface, and $ \Delta \rho $ is a threshold value typically set to a small increment (e.g., 0.03 $kg/m^3$) to capture the mixed layer’s depth relative to surface conditions.

To ensure a fair and consistent comparison of MLD across all models, we compute the MLD by sampling only the vertical levels that are common to all model configurations within the surface to 600 meters depth range. This approach eliminates discrepancies that could arise from differences in vertical resolution and guarantees that each model's MLD estimate is based on an identical set of depth points. By standardizing the vertical sampling, we ensure that variations in the derived MLD reflect true model performance rather than artifacts of differing vertical discretizations.
    
\begin{figure}[h!]
    \centering
    \includegraphics[width=\textwidth]{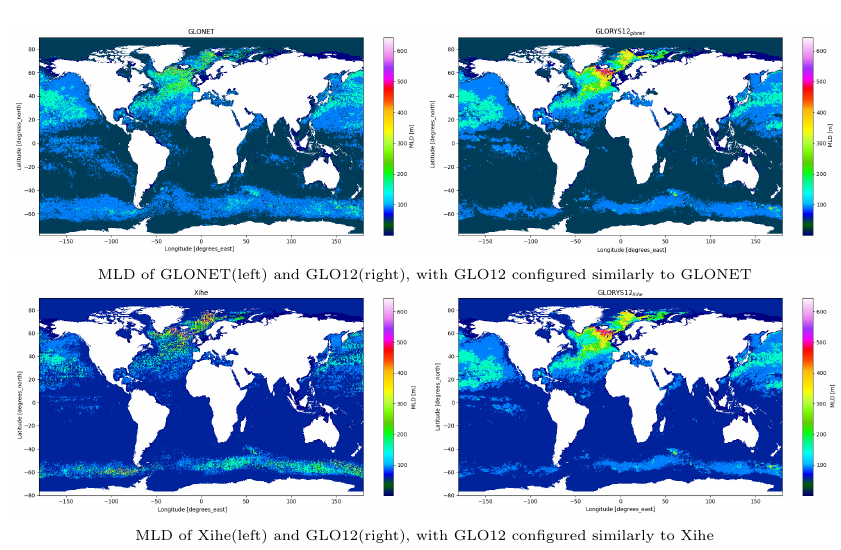}
    \caption{Examples of MLD reconstructed from GLONET and Xihe on 05-01-2024, along with those computed from GLORYS12 with two configurations replicating each time the vertical resolution of GLONET and Xihe.}
    \label{fig:mld}
\end{figure}

Figure~\ref{fig:mld} illustrates the derived MLD from GLONET, Xihe, and GLORYS12, with GLORYS12 again serving as a reference. GLONET’s MLD fields demonstrate a high degree of consistency with those of GLORYS12, accurately capturing large-scale spatial patterns of the mixed layer across the global ocean. In most regions, GLONET’s MLD estimates closely match the depth and extent observed in GLORYS12, indicating that GLONET is effectively preserving the dynamic relationship between temperature, salinity, and density. However, in the North Atlantic, where complex mixing processes and deep-water formation occur, GLONET’s MLD representation shows slight discrepancies compared to GLORYS12. This discrepancy suggests that while GLONET captures the broad structure of the mixed layer, there may be challenges in resolving finer-scale vertical mixing processes in regions with intense oceanic convection.

In contrast, Xihe's derived MLD fields exhibit pronounced noise, characterized by spatial artifacts that predominantly align along latitude and longitude directions. These patterns reflect architectural limitations within Xihe, where instabilities in the forecasted temperature and salinity fields lead to inconsistencies in the reconstructed density profiles. As a result, the MLD fields lack coherent structure and physical realism. The presence of such directional noise highlights challenges in Xihe’s ability to maintain the delicate balance between temperature and salinity, essential for accurate MLD estimation, and points to structural biases introduced by the model's design.

\paragraph{Statistical Analysis of Derived Quantities} In addition to visual comparisons, a statistical analysis is conducted to further quantify the accuracy of GLONET and Xihe in predicting derived quantities over a 10-days forecast horizon, averaged across the entire evaluation period from January to July 2024. Specifically, the geostrophic currents and mixed layer depth (MLD) are computed from GLONET's and Xihe's forecast outputs, comparing each to the GLORYS12 reanalysis, which serves as the reference standard. This approach mirrors the RMSE evaluations applied to primary variables, facilitating a consistent comparison of model performance across derived physical properties.

\begin{figure}[h!]
    \centering
    \includegraphics[width=1\textwidth]{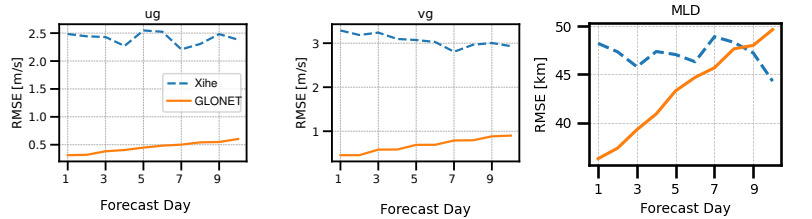}
    \caption{ Root Mean Square Error (RMSE) for the reconstructed geostrophic current and MLD for both GLONET, Xihe models. Calculations span from January to July 2024, using GLORYS12 as the reference. Geostrophic current and MLD are computed based on forecasts that are initialized weekly on Wednesdays, following the operational protocol.\label{fig:geo_mld}}
\end{figure}

The results (Figure \ref{fig:geo_mld}) show that GLONET maintains low RMSE values for both geostrophic currents and MLD throughout the forecast period, achieving close alignment with GLORYS12. This accuracy underscores GLONET’s capacity to produce forecasts that are not only quantitatively accurate but also dynamically consistent with observed ocean state dynamics.

In contrast, Xihe displays significantly higher RMSE values for the geostrophic currents, and a stable but slightly elevated MLD RMSE, indicating substantial deviations from GLORYS12. For geostrophic currents, the elevated RMSE suggests notable discrepancies in the SSH-derived fields, potentially driven by unresolved artifacts or inconsistencies. In the case of MLD forecasts, the relatively high RMSE is largely attributable to spatial artifacts, primarily aligned along latitude and longitude directions, which disrupt the coherence between temperature and salinity fields. This points to physical limitations in Xihe's ability to maintain consistency in its derived fields.

This statistical comparison highlights GLONET's ability to deliver dynamically consistent forecasts across derived variables, with low RMSE values that reflect its skill in reproducing complex ocean dynamics. On the other hand it also  reveals a limitation in using RMSE as the primary metric for evaluating data-driven models, as RMSE is not sensitive to small-scale inconsistencies or artifacts that affect the derived fields. RMSE measures point-wise accuracy and fails to capture the spatial continuity and coherence critical for processes like geostrophic flow derivation. This highlights the importance of process-oriented validation metrics. By deriving geostrophic currents from SSH,  a more comprehensive evaluation of each model's performance in preserving the integrity of physical fields is gained. While deriving the MLD offers insights into each model's ability to preserve physically consistent interactions between forecasted variables, particularly in capturing coherent and artifact-free stratification in the upper ocean.

\begin{figure}[h!]
    \centering
    \includegraphics[width=1\textwidth]{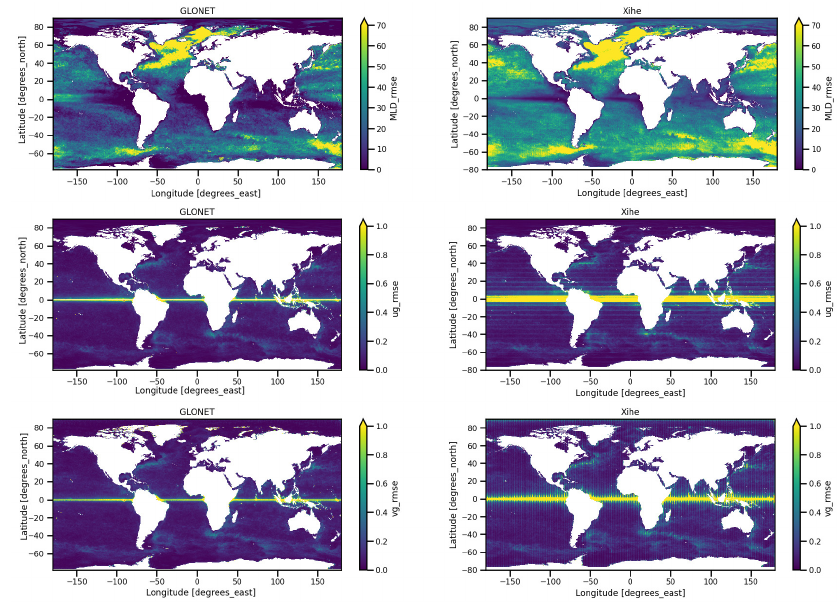}
    \caption{Spatial distribution of RMSE at the surface level for derived geostrophic currents and mixed layer depth (MLD) from GLONET and Xihe. The RMSE is computed with respect to GLORYS12 over the evaluation period from January to July 2024. These maps provide insight into the regional performance of each model in capturing derived physical properties, highlighting areas of stronger or weaker agreement with the reference reanalysis. Results are averaged over all lead times for weekly forecasts initialized on Wednesdays, adhering to the operational protocol.\label{fig:gmaprmse}}
\end{figure}

In addition to the analyses presented earlier, a complementary evaluation focusing on the spatial distribution of forecasting errors for derived quantities is provided, specifically geostrophic currents and mixed layer depth (MLD). This analysis aims to assess the regional variability of model performance, offering insights into how well GLONET and Xihe capture the spatial patterns of these derived physical properties.
The spatial RMSE maps, shown in Figure \ref{fig:gmaprmse}, reveal a stark contrast between the models. GLONET demonstrates a regional error distribution with coherent spatial patterns and relatively low errors across most regions. In contrast, Xihe shows significantly noisier error distributions, with high RMSE values scattered across all regions. This noisiness reflects inconsistencies in Xihe's forecasts, potentially linked to artifacts in the data and weaker coherence between variables such as temperature, salinity, and SSH. These results underscore GLONET's ability to preserve physical consistency in derived quantities.

\subsection{Diagnostic Variables}  
To complement the analysis of derived quantities, diagnostic variables that provide deeper insights into the intrinsic dynamics and energy-cascading properties of the forecasting models are explored. These diagnostics serve as critical tools for evaluating how well the models replicate fundamental ocean processes, including rotational dynamics and mesoscale variability.
In this context, the output of GLO12 is incorporated, not as a benchmark, but as a reference framework to evaluate the fidelity of neural network-based models like GLONET and Xihe in reproducing key oceanic features.
Particular emphasis is placed on vorticity and mesoscale eddies are investigated, both of which are directly tied to the dynamic integrity and physical realism of the modeled ocean states.  

\paragraph{Vorticity Analysis}  
In ocean dynamics, vorticity is a key diagnostic variable that characterizes the rotational motion of fluid parcels, offering insights into energy distribution, flow stability, and turbulence. Unlike derived quantities such as mixed layer depth (MLD) or geostrophic currents, which assess inter-variable coherence, vorticity delves into the system's internal dynamics and cascade-preserving capabilities. The vorticity, denoted as $\omega$, is calculated from the zonal ($u$) and meridional ($v$) components of the currents as follows:  

\begin{equation}  
\omega = \frac{\partial v}{\partial x} - \frac{\partial u}{\partial y},  
\end{equation}  

where $\frac{\partial v}{\partial x}$ represents the spatial derivative of the meridional velocity with respect to longitude, and $\frac{\partial u}{\partial y}$ is the spatial derivative of the zonal velocity with respect to latitude. This formulation captures the rotational characteristics of ocean currents, serving as a diagnostic for the conservation of small- and large-scale structures within the forecast.  

\begin{figure}[h!]
    \centering
        \includegraphics[width=\textwidth]{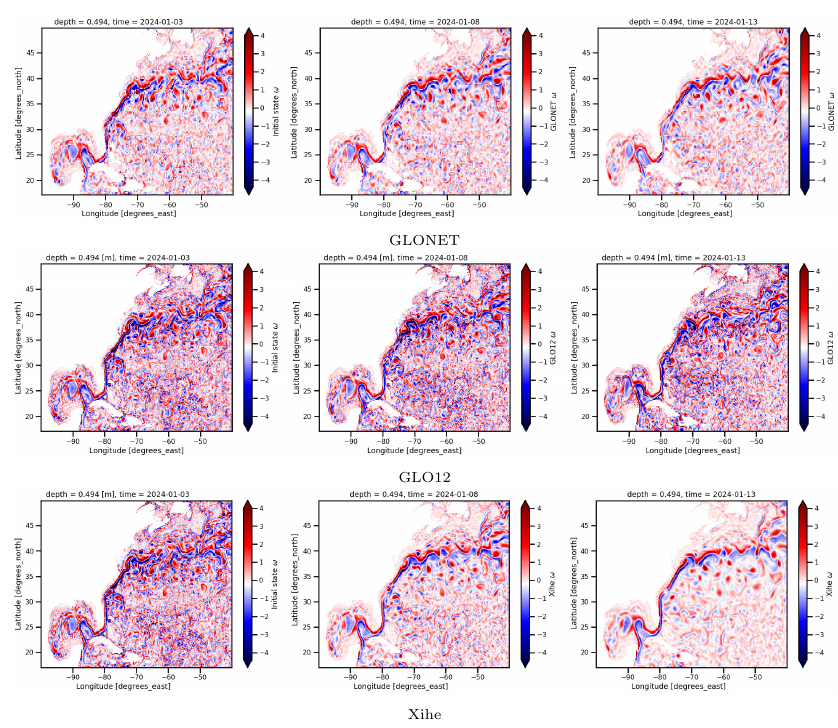}
    \caption{Evolution of vorticity over a 10-day forecast period for GLONET, GLO12, and Xihe, displayed over the Gulf Stream region. GLONET and GLO12 demonstrate stable energy-cascading and retain coherent rotational structures, whereas Xihe exhibits energy cascade attenuation, leading to progressive loss of small-scale features over time.}
    \label{fig:vorti}
\end{figure}

As shown in Figure~\ref{fig:vorti}, the vorticity fields for GLONET, GLO12, and Xihe over the Gulf Stream region reveal critical differences in dynamic fidelity. GLONET and GLO12 maintain robust coherence in vorticity across both small and large-scale structures, reflecting effective cascade-preserving capacities over the entire 10-day forecast period. This ability to preserve the energy transfer across scales is crucial for accurately capturing mesoscale phenomena and the energetic dynamics of ocean currents.  

In contrast, Xihe displays a significant energy cascade attenuation as the forecast progresses, with the fields appearing overly smooth by day 10. While large-scale rotational features persist, the loss of small-scale energy indicates an underlying instability or dissipation issue. This pattern resembles decaying turbulence, where energy cascades out of the system, undermining the representation of critical mesoscale dynamics. This behavior can be attributed to Xihe's architectural design, which employs independently trained models for each lead time. While this design stabilizes point-wise RMSE metrics, it neglects dynamic continuity, leading to the observed decay in fine-scale vorticity structures. This strategy might also neglects continuity across vertical levels as demonstrated in Figure~\ref{fig:vs_gs} in Appendix.

\begin{figure}[h!]  
    \centering  
    \includegraphics[width=1\textwidth]{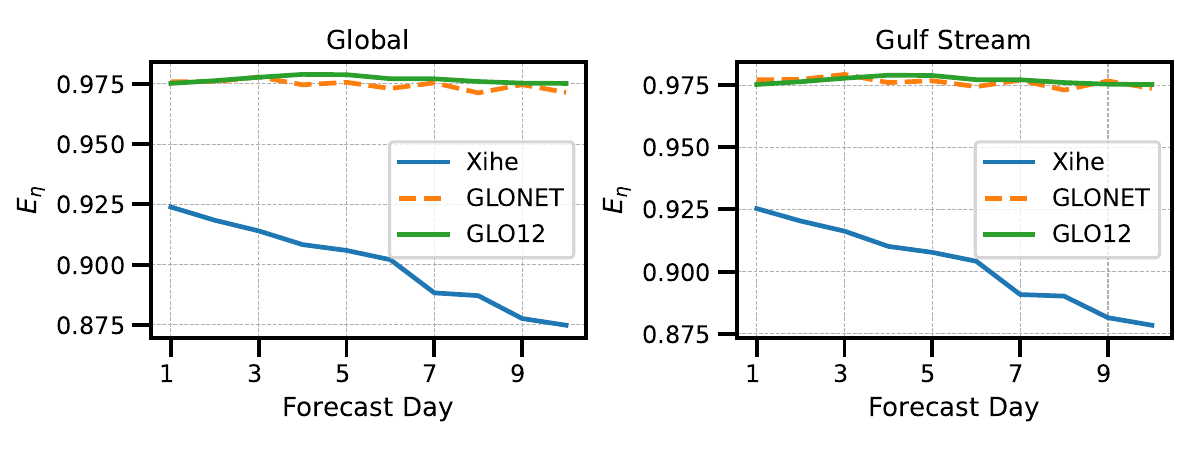}  
    \caption{Comparison of small-scale energy fraction over 10-days forecast period for Xihe, GLONET, and GLO12 models. The analysis was conducted from vorticity fields over the Gulf Stream (right) and the entire globe (left) over the evaluation period from January to July 2024. \label{fig:cascade}}  
\end{figure}

\paragraph{Energy Cascading Analysis:}
Beyond the visual analysis of cascade attenuation, where a progressive loss of small-scale features is observed over time, the energy dynamics is further investigated using a quantitative approach. To assess the retention of small-scale dynamics over time, the energy distribution, derived from the vorticity field, is analyzed across spatial scales. This way, the models' ability to preserve fine-grained dynamics critical for accurate ocean forecasting is evaluated.

The vorticity fields are analyzed in the spectral domain to determine their energy content at different spatial scales. Specifically, the vorticity field $\omega(x, y)$ is subjected to a two-dimensional discrete Fourier transform to obtain its spectral representation $\hat{\omega}(k_x, k_y)$, where $k_x$ and $k_y$ represent the zonal and meridional wavenumbers, respectively. The power spectrum is then computed as $|\hat{\omega}(k_x, k_y)|^2$, quantifying the energy associated with each spectral component.

To analyze the energy distribution in a more interpretable form, the two-dimensional power spectrum is radially averaged in wavenumber space, resulting in the one-dimensional energy spectrum $E(k)$, where $k = \sqrt{k_x^2 + k_y^2}$. This spectrum represents the energy density as a function of total wavenumber, enabling the characterization of energy across spatial scales.

Scale-integrated energy content is defined as the sum of $E(k)$ over all wavenumbers $k < k_c$, where the cutoff wavenumber $k_c$ corresponds to a physical length scale $L_c$ (e.g., $L_c = 100 \, \text{km}$). Note that a cutoff of $L_c = 100 \text{km}$ includes all spectral energy from 100 km and above, while larger cutoff values (e.g., 500 or 1000 km) exclude progressively more small-scale energy. The total energy is computed by summing $E(k)$ across all wavenumbers, and the fraction of energy associated with $L_c$ cutoff is then defined as:

\begin{equation}
E_{\eta} = \frac{\sum\limits_{k < k_c} E(k)}{\sum\limits_{k } E(k)}
\end{equation}

This dimensionless fraction quantifies the proportion of total energy integrated over spatial scales bounded by $Lc$ relative to the entire spectrum. The goal is not to assess the absolute energy retained but rather whether the spectral shape, particularly at small scales, is preserved across lead times, ideally showing a consistent profile similar to the initial condition.

By examining the scale-integrated energy fraction over the 10 days forecast horizon, the ability of each model to maintain the energy cascade characteristic of dynamic fidelity is evaluated. This analysis is conducted for 10 days forecasts spanning the period from January to July 2024, covering both global domains and the Gulf Stream region. The results consistently revealed that both GLONET and GLO12 maintained stable small-scale energy fractions throughout the forecast period, demonstrating their capacity to preserve fine-scale features and sustain energy consistency. Notably, the stability of these fractions was observed across both the global domain and the Gulf Stream region, underscoring the robustness of GLONET and GLO12 in preserving small-scale dynamics.

Conversely, Xihe exhibited a marked decline in the small-scale energy fraction over the forecast horizon, indicative of its inability to sustain the energy cascade. This pattern, consistent across both the global domain and the Gulf Stream region, aligns with the observed decay of vorticity structures in Xihe forecasts. To further isolate whether this decay is related specifically to the loss of small-scale energy, we repeated the analysis using different cutoff scales ($L_c = 200$, 500, and 1000 km), as presented in the appendix. Additionally, we conducted a power spectral density (PSD) analysis of the vorticity field to more precisely attribute energy loss to specific spatial scales. These extended analyses confirm that Xihe's energy decay spans a wide range of scales, highlighting broader challenges in maintaining dynamical coherence and preserving the intricate coupling between scales essential for accurate ocean forecasting.

\paragraph{Mesoscale Eddies}  
Mesoscale eddies, integral to ocean dynamics, were analyzed to assess the models' capacity to represent coherent structures. Eddies were identified using an Eulerian approach, based on detecting closed SSH contours with a monotonic increase or decrease towards a central extremum. To ensure consistency across models, no smoothing was applied to the SSH fields, and a convexity criterion of at least 0.9 was imposed to confirm eddies coherence.  
Over the 10-day forecast horizon, no significant temporal evolution in eddy distribution is expected, therefore this analysis did not involve tracking individual eddies. Instead, it focuses on detecting spatial patterns at each lead time.  

\begin{figure}[h!]
    \centering
        \includegraphics[width=\textwidth]{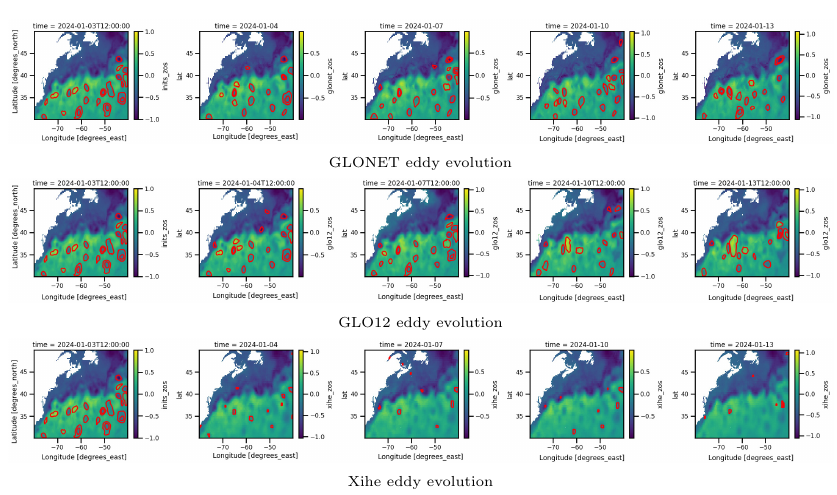}
    \caption{Evolution of eddies over a 10-day forecast period for GLONET, GLO12, and Xihe, displayed over the Gulf Stream region. The first column represents the eddies extracted from the initial state inputted to all the models.}
    \label{fig:eddies}
\end{figure}

The analysis shows that GLONET and GLO12 perform comparably well, accurately capturing the spatial distribution of mesoscale eddies over the 10-day forecast period. Figure ~\ref{fig:eddies} illustrates the regional distribution of detected eddies, revealing that both GLONET and GLO12 align closely with the expected patterns of mesoscale variability. This consistency underscores their ability to maintain the SSH gradients necessary for eddy detection and to preserve dynamic integrity across lead times.  

Conversely, Xihe exhibits significant shortcomings, frequently failing to detect eddies or misrepresenting their spatial structure. This deficiency highlights Xihe's difficulty in maintaining coherent SSH gradients, which is essential for identifying mesoscale features. 
These results emphasize the critical role of dynamic consistency in accurately representing mesoscale processes. Both vorticity and mesoscale eddies serve as diagnostic variables that reveal how well forecasting models replicate the fine-scale and energetic dynamics of ocean circulation.

\subsubsection{Lagrangian analysis}

In this subsection, the process-oriented accuracy of GLONET is assessed by analyzing Lagrangian trajectories over a forecast horizon of 10 days, using daily forecast outputs from January to July 2024.  Lagrangian drift analysis offers insight into a model's ability to capture the advection of ocean particles over time, which is critical for applications involving transport processes such as pollutant dispersion, larval connectivity, and passive tracer dynamics. By simulating the motion of synthetic particles advected by model-predicted velocity fields, we assess whether the flow structures are coherent and physically realistic.

Let's consider the ocean currents field:

\begin{equation}
    \mathbf{v}(\mathbf{x},t), \quad \mathbf{x} \in \mathbb{R}^2, \quad t\in [t_0, t_{f}] 
\end{equation}

and its associated ordinary differential equation:

\begin{equation}
    \dot{\mathbf{x}}= \mathbf{v}(\mathbf{x},t), \quad \mathbf{x} \in \mathbb{R}^2, \quad t\in [t_0, t_{f}] 
\end{equation}

where $\mathbf{v}$  the U and V components of ocean currents, defined on a possibly time-dependent spatial domain $\mathcal{U}(t) \in \mathbb{R}^2 \times [t_0, t_{f}]$.

Lagrangian trajectories are defined as: 
\begin{equation}
    \mathbf{x}(t_f, t_0, \mathbf{x}_0)= \mathbf{x}_0+\int_{t_0}^{t_f} \mathbf{v}(\mathbf{x}(\tau),\tau) \, d\tau
\end{equation}

\begin{figure}[h!]
    \centering
    \includegraphics[width=\textwidth]{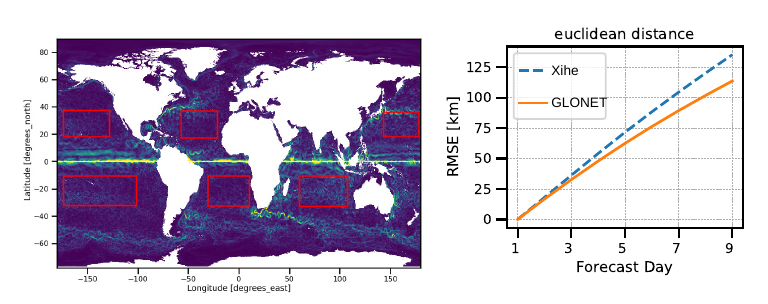}
    \caption{ (Left) Red boxes highlight the areas where particles were initiated. (Right) Divergence of  GLONET and Xihe's Lagrangian trajectories from those of GLORYS12. Trajectories reconstruction is done based on the weekly forecasts span from from January to July 2024. \label{fig:lag}}
\end{figure}

Following the above, Lagrangian trajectories of the different models are computed by setting up $\mathbf{v}$ each time to ocean current forecasts from GLONET, GLO12 and Xihe, with the flow integration being carried out following a fourth-order Runge-Kutta method with variable step-size.
For each model, particles initialized at identical locations across the red boxes of Figure \ref{fig:lag} are tracked, and  Euclidean distances between the modeled trajectories and those from GLORYS12 are calculated by setting up $t_0$ to the first lead-day forecast and $t_f$ to the $10th$ lead-day. This Euclidean distance metric provides a quantitative assessment of each model’s accuracy in reproducing the physical transport patterns present in the reanalysis data.

As shown in Figure \ref{fig:lag}, GLONET consistently outperforms Xihe, maintaining closer proximity to the GLORYS12-derived trajectories. At the 10-day lead time, GLONET reduces the Euclidean distance error by approximately 25 km. This improvement reflects GLONET’s enhanced capacity to capture accurate flow dynamics and oceanic transport structures over extended forecast horizons.

\paragraph{Comparison of Geostrophic and Full Lagrangian Trajectories}

To further evaluate the dynamical coherence within each model, Lagrangian trajectories are reconstructed using both geostrophic currents (derived from SSH) and full currents for GLONET, GLO12, and Xihe. By comparing the Euclidean distances between these geostrophic and full trajectories, the degree of alignment between SSH and current fields are assessed, offering insight into each model’s capability to represent cohesive, physically consistent flow dynamics.

\begin{figure}[h!]
    \centering
    \includegraphics[width=0.5\textwidth]{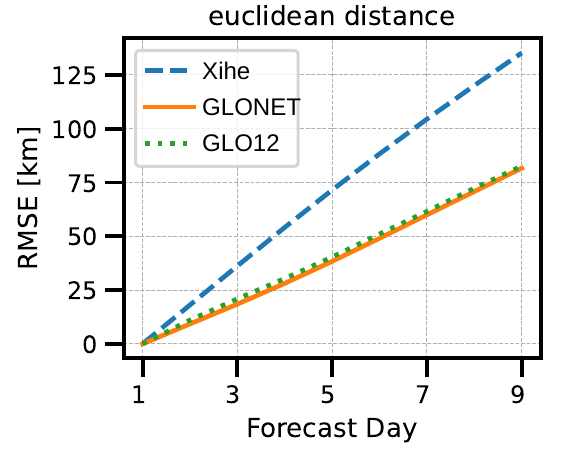}
    \caption{ Divergence of geostrophic Lagrangian trajectories from those computed using full current each time for GLO12, GLONET and Xihe. Trajectories calculations is done based on the weekly forecasts of ocean currents span from January to July 2024 along with geostrophic currents approximated from the forecasted SSH. \label{fig:glag}}
\end{figure}

The results, illustrated in Figure ~\ref{fig:glag}, show that GLONET and GLO12 exhibit consistent RMSE evolution over the 10-day forecast period, with a stable and generally smaller Euclidean distance between geostrophic and full trajectories. This close alignment indicates that both models maintain a high degree of coherence between SSH and current predictions, reflecting an ability to generate forecasts with internally consistent surface and total flow fields.
Conversely, Xihe demonstrates a rapid divergence between geostrophic and full current-based trajectories, evident in the increasing Euclidean distance over time. This discrepancy suggests a lack of coherence between Xihe's SSH and current fields, likely due to artifacts that surface in derived geostrophic currents. The divergence highlights Xihe’s limitations in producing dynamically consistent outputs across interconnected variables.
 \section{Discussion and Summary}  
\label{sec:diss_sum}

A comprehensive evaluation of GLONET, a novel data-driven ocean forecasting model, is conducted by comparing its performance against two systems: GLO12, a physics-driven operational model, and Xihe, an AI-based forecasting approach. The assessment encompassed multiple dimensions, including traditional point-wise error metrics, process-oriented analyses, diagnostic variables, derived physical quantities, and observation-based evaluations using the IV-TT Class 4 Framework. This multidimensional perspective highlights GLONET’s strengths and limitations relative to established approaches.

The CLASS4 framework provided an independent evaluation of the systems’ performance by comparing model outputs directly against in situ and satellite observations. Analyses of temperature and salinity in the 5–100 m layer, SST from drifting buoys, and SLA from satellite altimeters revealed that GLONET consistently outperformed Xihe across most metrics and forecast lead times. In particular, GLONET demonstrated superior accuracy in SLA forecasts, aligning closely with GLO12 for intermediate lead times (days 5–9). However, GLO12 displayed an advantage in SST forecasting, especially in tropical regions, underscoring the challenges of accurately resolving surface variables in data-driven models. These CLASS4-based evaluations emphasized GLONET’s capacity to produce forecasts comparable to operational standards while highlighting areas for further refinement.

A key distinction between the three models, GLONET, Xihe, and GLO12 lies in their underlying methodologies. GLONET is a data-driven model trained on historical outputs of physical systems, without direct assimilation of real-time observations or governing physical laws. This leads to its strong performance in large-scale dynamics and ocean currents, but also to a noticeable gap in surface predictions such as SST, where GLO12 benefits from direct assimilation of satellite observations. Xihe, like GLONET, is data-driven, but with a distinct training strategy that uses separate models for each forecast lead time, enabling stable performance across lead times. However, this comes at the cost of dynamic continuity, as demonstrated by Xihe's weaker performance in preserving derived quantities like geostrophic currents and mesoscale eddies. GLO12, as a physics-based model, excels in surface variables like SST, due to its reliance on both physical equations and real-time data assimilation, making it the most accurate for tropical surface temperature predictions.

Point-wise evaluations, conducted using GLORYS12 as the reference dataset, assessed the models’ ability to generate forecasts consistent with the data used during training. GLONET exhibited strong performance in predicting ocean currents and comparable accuracy in temperature and salinity forecasts, reflecting its capability to capture large-scale ocean dynamics. GLONET also experienced a gradual decline in forecast accuracy with increasing lead times, characteristic of autoregressive models, whereas Xihe maintained stable error levels due to its lead-time-specific training strategy. This stability, however, came at the expense of inter-variable consistency and dynamic coherence.

Extending beyond point-wise assessments, process-oriented analyses revealed critical aspects of model performance that are often overlooked in traditional metrics. For instance, GLONET demonstrated a strong ability to derive geostrophic currents from forecasted SSH, maintaining physical consistency and dynamic coherence akin to GLORYS12. Xihe, in contrast, exhibited significant noise, exposing artifacts in its derived quantities, stemming from its RMSE-focused training strategy. Similarly, GLONET’s mixed layer depth (MLD) estimates closely matched GLORYS12, except for localized discrepancies in regions such as the North Atlantic, where finer-scale adjustments to temperature and salinity stratification might be required.
These results highlight GLONET's ability to maintain consistency between temperature and salinity, a critical requirement for accurately representing ocean stratification. Xihe's MLD predictions, on the other hand, were plagued by noise and inconsistencies, further emphasizing the need for enhanced regularization in AI-driven approaches.  

Diagnostic analyses further evaluated the models’ fidelity in preserving dynamic consistency and fine-scale structures. Vorticity provided a direct measure of the rotational motion within the models, serving as a diagnostic for energy cascading and dynamic coherence. Both GLONET and GLO12 successfully preserved vorticity structures over the Gulf Stream region, maintaining coherent small and large-scale features over the 10-day forecast period. In contrast, Xihe displayed a progressive loss of small-scale features, transitioning toward a state resembling decaying turbulence by day 10. This behavior stems from Xihe's architectural design, which uses independently trained models for each lead time, neglecting dynamic continuity and leading to the observed degradation.  

Lagrangian trajectory analyses provided additional insights into dynamic accuracy. Particle tracking over a 10-day forecast horizon highlighted GLONET’s superior ability to maintain coherence across time and space, outperforming Xihe by approximately 25 km in trajectory accuracy. Both GLONET and GLO12 exhibited consistent alignment between geostrophic and full currents, reflecting a strong coupling between SSH and surface velocity fields. Xihe, however, displayed rapid divergence between these trajectory types, revealing weaknesses in its physical coupling.

Mesoscale eddies, critical to ocean energy transport and mixing, were identified using closed SSH contours with a convexity criterion of 0.9, ensuring fair comparison across models without smoothing the SSH fields. GLONET and GLO12 demonstrated strong consistency in eddy detection, capturing the spatial distribution of mesoscale features accurately over the forecast horizon.  In contrast, Xihe often failed to detect eddies or misrepresented their structure, reflecting deficiencies in maintaining the SSH gradients necessary for identifying coherent eddy features. These results emphasize the importance of dynamic consistency and structural integrity for accurately resolving mesoscale phenomena.  

To complement these analyses, spatial RMSE maps of surface variables and derived quantities provided a regional perspective on model performance. GLONET and GLO12 exhibited nearly identical spatial error distributions, suggesting that both models are influenced by similar physical challenges, such as resolving western boundary currents and equatorial dynamics. Xihe, however, showed a more erratic spatial error pattern, characterized by pervasive noise across derived quantities. This disparity underscores the value of incorporating physical constraints into data-driven models to ensure spatial coherence and reduce artifacts in regions with high variability.

In summary, this study underscores the potential of GLONET as a next-generation ocean forecasting model that effectively bridges the gap between traditional physics-based systems and purely data-driven approaches. By capturing key ocean dynamics, ensuring inter-variable consistency, and producing forecasts competitive with operational standards, GLONET demonstrates its capacity to serve as a robust forecasting tool. Nevertheless, localized discrepancies and the inherent limitations of purely data-driven methodologies highlight the need for further refinements. This work also illustrates the necessity of comprehensive evaluation frameworks in ocean forecasting. Traditional metrics like RMSE, while informative, are insufficient to fully capture the dynamical fidelity and inter-variable coherence required for reliable forecasts. Process-oriented assessments, diagnostic variables and derived quantity analyses provide crucial insights into data-driven models performance, ensuring that forecasts not only minimize errors but also respect the fundamental dynamics of the ocean system. Together, these findings pave the way for advancing data-driven ocean forecasting and offer a foundation for future improvements in methodology and application.

\section*{Open Research Section}
\label{sec:data_avai}

GLONET $[1/4^{\circ}]$ experimental daily forecasts are available on EDITO
platform at \url{https://glonet.lab.dive.edito.eu/}.     

GLORYS12 and GLO12 products are available on the Copernicus Marine Service
at \url{https://data.marine.copernicus.eu/}.
 \section*{Acknowledgements}
\label{sec:ack}

We are thankful to X. Wang et al. for providing access to their pre-trained model. \section*{Author Contribution Statement }
\label{sec:contribution}
A.E conceived the study, developed the model, conducted the simulations/experiments, and wrote the original draft of the manuscript.
Q.G built the pre-operational pipeline for GLONET's integration.
C.R conducted the Class-4 evaluation.
All authors discussed the results and contributed to the final manuscript. \bibliographystyle{apalike}
\bibliography{ml.bib,ocean.bib}
\section*{Appendix}
\label{sec:appen}

\subsection*{Vertical sections evaluation}

\begin{figure}[h!]
    \centering
    \includegraphics[width=1\textwidth]{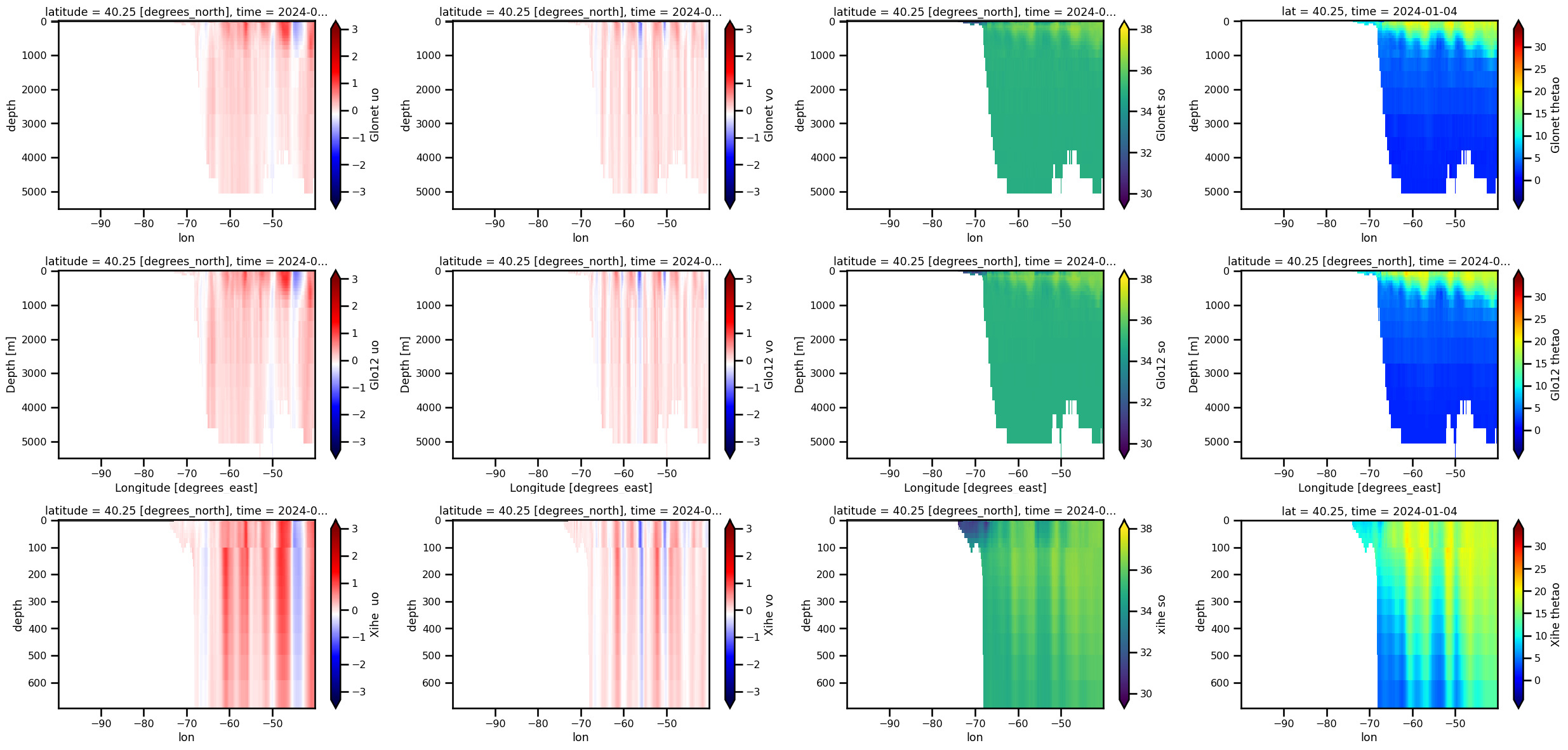}
    \caption{Vertical sections of the Gulf Stream for 3D variables (U and V currents, temperature, and salinity) from GLO12, GLONET, and Xihe on the same date. Notably, Xihe shows a pronounced 100 m depth shift, particularly evident in the zonal (U) and meridional (V) current components.  \label{fig:vs_gs}}
\end{figure}

The vertical sections of temperature, salinity, and ocean currents, depicted in Figure~\ref{fig:vs_gs}, provide a detailed comparative assessment of the models in the Gulf Stream region at a latitude of $40.25^{\circ}N$. Both GLONET and GLO12 demonstrate smooth and continuous vertical distributions, highlighting their ability to accurately capture vertical coherence and dynamic consistency across the water column. These results reflect their robust handling of vertical gradients and fine-scale patterns essential for representing ocean stratification and circulation dynamics.
In contrast, Xihe exhibits a distinct discontinuity around the 100-meter depth, particularly evident in the ocean current profiles. This artifact arises from its architectural design, which employs two independent models: one optimized for depths between the surface and 100 meters, and another for depths from 100 to 600 meters. While temperature and salinity fields are less impacted, showing minor inconsistencies at the model transition depth, the pronounced discontinuity in ocean currents highlights the challenges associated with maintaining smooth transitions across depth ranges in data-driven frameworks that rely on segmented modeling.

\subsection*{Energy cascade and power spectrum analysis}

\begin{figure}[H]
    \centering
    \includegraphics[width=1\textwidth]{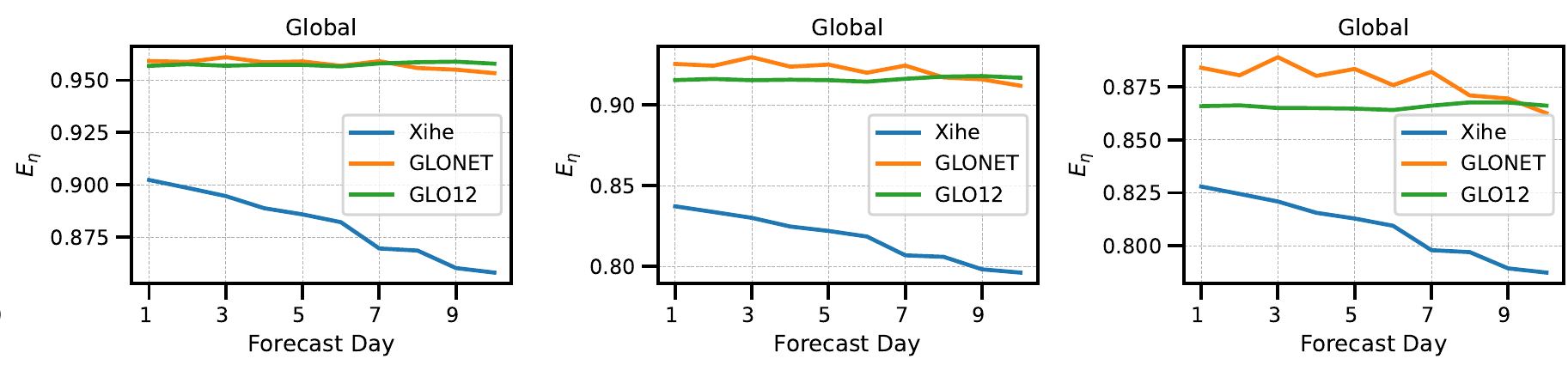}
    \caption{Comparison of small-scale energy fraction over the 10-day forecast period for Xihe, GLONET, and GLO12 models. The analysis is based on vorticity fields over the global ocean and spans the evaluation period from January to July 2024. From left to right, panels correspond to cutoff scales of 200 km, 500 km, and 1000 km, respectively.  \label{fig:cascad2}}
\end{figure}

To complement the main analysis of small-scale energy retention, additional evaluations were performed using alternative cutoff wavelengths ($L_c = 200$, 500, and 1000~km) to test the sensitivity of the results to spectral thresholding (see Figure~\ref{fig:cascad2}). These supplementary analyses corroborated the primary findings: both GLONET and GLO12 consistently maintained stable $Lc$-integrated scale energy fractions over the 10-day forecast horizon, indicating a robust capacity to preserve the energy cascade and fine-scale dynamics. Interestingly, at coarser cutoffs (particularly $L_c = 500$ and $L_c = 1000$~km), GLONET tends to retain slightly higher energy levels than GLO12, suggesting a marginally more energetic representation at intermediate scales. In contrast, Xihe continued to exhibit a marked decline in energy across all cutoff levels, reinforcing its limitations in sustaining dynamical coherence and multiscale coupling.

\begin{figure}[H]
    \centering
    \includegraphics[width=1\textwidth]{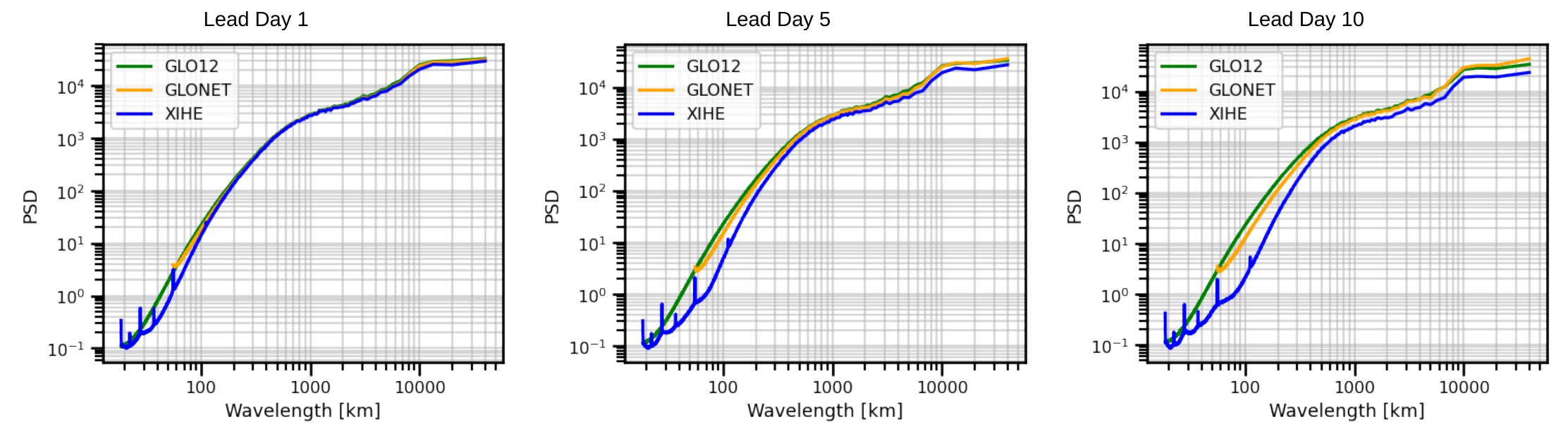}
    \caption{Power spectral density averaged from January to July 2024 for the entire globe. The analysis is based on vorticity fields and illustrates the distribution of energy across spatial scales at different forecast lead times (day 1, 5, and 10).  \label{fig:psd}}
\end{figure}

To further contextualize the small-scale energy retention analysis, we perform a power spectral density (PSD) evaluation of surface vorticity for GLONET, Xihe, and GLO12 at lead times of 1, 5, and 10 days (see Figure~\ref{fig:psd}). This spectral decomposition provides a more detailed view of the energy distribution across spatial scales, offering a complementary perspective to the scale-integrated energy fractions. The results reveal that Xihe systematically loses energy across all wavelengths with increasing lead time, with the depletion being most pronounced at smaller scales (below 200~km). Additionally, Xihe exhibits anomalously high spectral peaks at these finer scales, indicative of noise or unphysical variance accumulating in the absence of dynamical constraints. In contrast, GLO12 maintains a consistent spectral shape across lead times, indicative of stable multiscale dynamics. GLONET displays only mild spectral evolution: while it shows a modest reduction in energy at the smallest scales relative to day-1 forecasts, it retains higher energy at larger scales compared to GLO12. This slight redistribution suggests that GLONET remains energetically coherent over time, albeit with a subtle shift in its spectral energy balance toward broader features.

\subsection*{Vertical Analysis: GLONET vs GLO12}

\begin{figure}[H]
    \centering
    \includegraphics[width=1\textwidth]{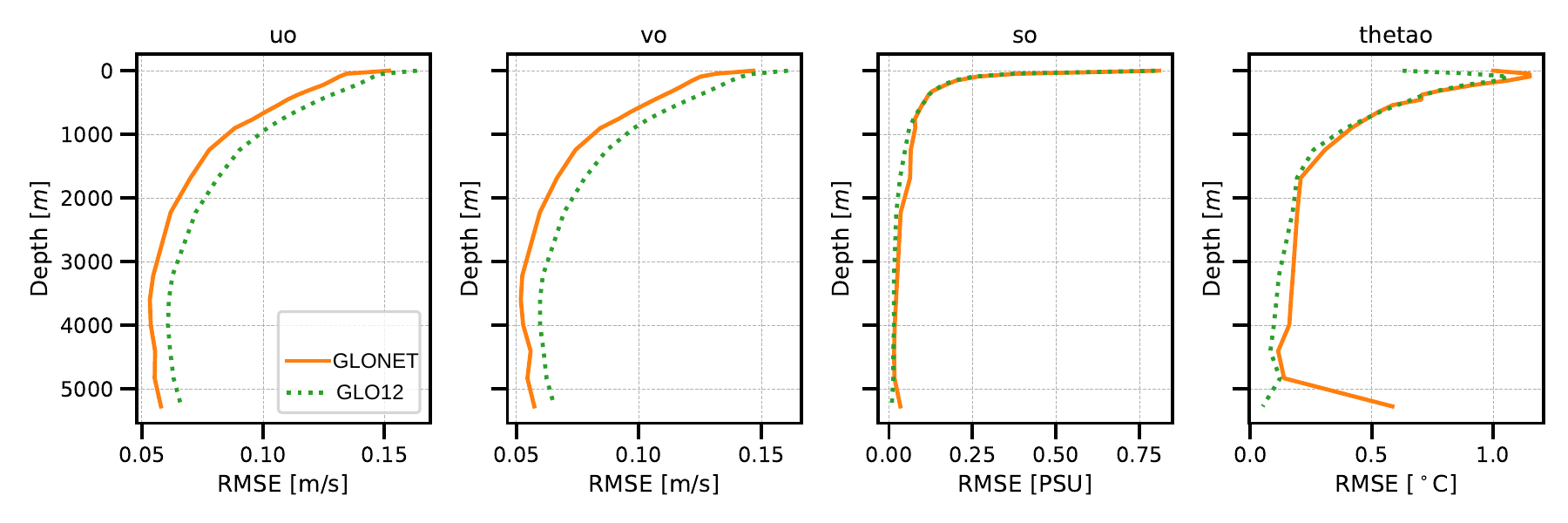}
    \caption{RMSE computed at each depth level and averaged over all lead times for 3D variables (U and V currents, temperature, and salinity) for the GLONET (orange line) and GLO12 (green line) models, covering January to July 2024. GLORYS12 serves as the reference, with 10-day forecasts initialized weekly on Wednesdays from a nowcast analysis performed with GLO12 seven days behind real-time.\label{fig:depthrmse2}}
\end{figure}

Building upon the depth-resolved evaluation presented in earlier (Figure~\ref{fig:depthrmse}), a dedicated comparison between GLONET and GLO12, was conducted using GLORYS12 as the reference (see Figure~\ref{fig:depthrmse2}). This comparison enables a more balanced assessment of their respective skills across the vertical column. Overall, GLONET demonstrates improved performance in forecasting ocean currents throughout the water column, reflecting its ability to capture and propagate dynamical features more effectively. However, for hydrographic variables such as temperature and salinity, GLONET exhibits slightly lower accuracy than GLO12, particularly at greater depths. These findings suggest that while GLONET offers enhanced dynamical fidelity, it remains more challenged in representing the thermohaline structure of the deep ocean, where GLO12's physically constrained formulation provides a modest advantage.

\subsection*{Architectural Rationale and Component-Level Considerations}
\paragraph*{Spectral and Spatial Feature Modeling}
The spatial architecture of GLONET integrates both spectral and spatial representations to effectively model multi-scale ocean dynamics. The Fourier Neural Operator (FNO) is employed to capture large-scale, low-frequency components by leveraging global receptive fields and efficient spectral parameterizations. While FNO excels at modeling smooth, basin-wide structures such as gyres and planetary waves, its inherent spectral bias limits its capacity to resolve high-frequency, localized phenomena. To mitigate this, a convolutional neural network (CNN) module is incorporated in parallel. The CNN, characterized by localized convolutional kernels, complements the FNO by enhancing sensitivity to sharp gradients and fine-scale structures, including submesoscale eddies and fronts. This dual-branch configuration enables GLONET to capture the full range of spatial variability present in ocean fields.

\paragraph*{Temporal Encoding and Implicit Forcing Representation}
Temporal information is encoded through a dedicated operator acting on two consecutive states of the ocean. This module is designed to extract the underlying dynamical evolution, thereby implicitly capturing the effects of surface forcing and transient tendencies without the need for explicit atmospheric inputs. By conditioning on both $X_{t-1}$ and $X_t$, the temporal encoder learns representations that embed short-term forcings into the model state, which is particularly effective for short-range forecasts where the predictive horizon does not critically depend on long-term external drivers.

\paragraph*{Multi-Scale Fusion via Latent-Space Composition}
To facilitate integration of representations extracted at disparate spatial scales, GLONET employs an encoder-decoder framework. Rather than aggregating the outputs of the FNO and CNN modules in the physical domain, where representational incompatibilities may impair learning, the model maps both outputs into a shared latent space. This design enables more coherent fusion by aligning features at an abstract representational level, thereby improving the model's ability to fuse multi-scale information in a physically consistent and data-efficient manner. The decoder then reconstructs the forecasted ocean state from this latent representation, completing the operator mapping from initial condition to future state.

\begin{table}[h]
\centering
\caption{List of acronyms.}
\label{tab:abbreviations}
\begin{tabular}{ll}
\hline
\textbf{Acronyms } & \textbf{Definition} \\
\hline
ACC & Antarctic Circumpolar Current\\
AI & Artificial Intelligence\\ 
AIFS & Artificial Intelligence/Integrated Forecasting System\\
CNN & Convolutional Neural Network\\
EDITO & European Digital Twin Ocean platform EDITO\\
FNO & Fourier neural operators\\
GDP & Global Drifter Program\\
GLO12 & global ocean analysis and forecast system\\
GLORYS12 & Global Ocean Reanalysis and Simulation at $1/12^\circ$ resolution (Version 12)\\
GPU & Graphics Processing Unit\\
HPC & High Performance Computing\\
IV-TT & Class IV Task Team\\
MAE & Mean Absolute Error\\
ML & Machine Learning\\
MLD & Mixed Layer Depth\\
NWP & Numerical Weather prediction\\
RMSE & Root Mean Square Error\\
RMSD & Root Mean Square Difference\\
SLA & Sea Level Anomaly\\
SSH & Sea Surface Height\\
SST & Sea Surface Temperature\\
\hline
\end{tabular}
\end{table}

\end{document}